\documentclass[aps,preprint,showpacs,preprintnumbers,amsmath,amssymb,footi
nbib]{revtex4-1}
\usepackage[colorlinks=true, pdfstartview=FitV, linkcolor=red, citecolor=blue, urlcolor=blue]{hyperref}
\usepackage{graphicx}
\usepackage{bm}

\newcommand{\bdy}{{\bold Y}}
\newcommand\beq{\begin{equation}}
\newcommand\eeq{\end{equation}}

\usepackage{amsmath} \begin{document}

\preprint{RBRC-993} \preprint{BNL-99061-2013-JA} 
\title{The Roberge-Weiss transition and 't Hooft loops}

\author{Kouji Kashiwa} 
\email[]{kashiwa@ribf.riken.jp}
\affiliation{RIKEN/BNL, Brookhaven National Laboratory, Upton, NY 11973}

\author{Robert D. Pisarski} 
\email[]{pisarski@bnl.gov} 
\affiliation{
Department of Physics, Brookhaven National Laboratory, Upton, NY 11973}
\affiliation{RIKEN/BNL, Brookhaven National Laboratory, Upton, NY 11973}

\begin{abstract} 
Roberge and Weiss showed that for $SU(N)$ gauge theories, 
phase transitions occur in the presence of an imaginary quark
chemical potential.
We show that at asymptotically high temperature, where
the phase transition is of first order,
that even with dynamical quarks 't Hooft loops of arbitrary $Z(N)$ charge
are well defined at the phase boundary.  
To leading order in weak coupling, 
the 't Hooft loop satisfies Casimir scaling
in the pure glue theory, but not with quarks.
Because the chemical
potential is imaginary, typically the interaction measure is negative on one
side of the phase transition. 
Using a matrix model to model
the deconfining phase transition, we compute the phase diagram for
heavy quarks, in the plane of temperature and imaginary chemical potential.
In general we find intersecting lines of first order transitions.
Using a modified Polyakov loop which is Roberge-Weiss
symmetric, we suggest that 
the interface tension is related to the 't Hooft loop only at high temperature,
where the imaginary part of this Polyakov loop,
and not the real part, is discontinuous across the phase boundary.
\end{abstract}

 \maketitle


\section{Introduction}

Understanding the nature of the phase diagram of Quantum Chromodynamics
(QCD) is one of the outstanding problems in nuclear physics.
At zero chemical potential, the theory can be studied by numerical
simulations on the lattice.  Generally this is not possible at
nonzero chemical potential because of the sign problem \cite{Forcrand:2009}.

One way to trying to understand the theory for nonzero chemical potential
is to consider a chemical potential which
is purely imaginary: then there is no sign problem, and numerical simulations
are possible.  
Although an imaginary chemical potential is not directly
physical, results can be related to those for
real chemical potential using
Fourier and Laplace transformations~\cite{Roberge:1986}.  

The pure gauge theory is invariant under a global symmetry of $Z(N)$,
but this is violated when dynamical quarks are present.
Roberge and Weiss showed, however, that in the presence of $\phi$,
an imaginary quark chemical potential, that the theory acquires
a global $Z(N)$ symmetry, under which
$\phi \rightarrow \phi + 1/N$ \cite{Roberge:1986}.  
They also showed that at high temperature,
there is a first order transition at constant
temperature, as $\phi$ is varied across $\phi_{RW} = 1/(2 N)$.
We generalize this to $Z(N)$ transformations of charge $k$,
and find a first order transition when $\phi_{RW} = k/(2 N)$.
As the temperature is lowered, this single line of transitions
can split into two lines at $\phi \neq \phi_{RW}$; these can be
a true phase transition, of first or second order, or just crossover.
For a transition which occurs when $\phi \neq \phi_{RW}$,
the Roberge-Weiss transition mixes either with the transitions for
deconfinement or for the restoration of chiral symmetry.
This has been studied by numerical simulations
on the lattice 
\cite{Alford:1998sd, *Hart:2000ef, *deForcrand:2002ci, *D'Elia:2002gd, *D'Elia:2004at, *Chen:2004tb, *deForcrand:2006pv, *Wu:2006su, *D'Elia:2007ke, *deForcrand:2008vr, *D'Elia:2009tm, *D'Elia:2009qz, *deForcrand:2010he, *Cea:2012ev, Bonati:2011a, *Bonati:2011b, *Nagata:2011yf, *Wu:2013bfa, Fromm:2012yg}, 
in effective theories \cite{Bluhm:2007cp, *Braun:2007bx, *Sakai:2008um, *Sakai:2008ga, *Kashiwa:2008bq, *Kouno:2009bm, *Kashiwa:2009ki, *Sakai:2009dv, *Sakai:2009vb, *Braun:2009gm, *Nam:2009nn, *Matsumoto:2010vw, *Kouno:2011vu, *Kouno:2011zu, *Sasaki:2011, *Morita:2011eu, *Morita:2011jva, *Pagura:2011rt, *Kashiwa:2012xm, *Fister:2013bh},
and in holographic models \cite{Aarts:2010ky, *Rafferty:2011hd}.

In this paper we show that Roberge-Weiss phase transitions have
an unexpected connection with pure glue theories.
Consider $SU(N)$ gauge theories without dynamical quarks.
Running the Wilson loop in the temporal direction gives the Polyakov loop.
This is an order parameter for 
$Z(N)$ electric charge, whose expectation value 
vanishes in the confined phase and is nonzero
in the deconfined phase.  Similarly, the two point function of Polyakov
loops exhibits an area law in the confined phase.

The response of the theory to $Z(N)$ magnetic charge
is given by the 't Hooft loop
\cite{'tHooft:1979uj,KorthalsAltes:1999xb, *KorthalsAltes:2000gs, deForcrand:2001nd, *deForcrand:2000fi, *deForcrand:2005pb, Lucini:2012gg}.
Its behavior is converse to that of the Wilson or Polyakov loops,
in that the 't Hooft loop exhibits an area law only in the deconfined phase.
The string tension for the area law 
of the 't Hooft loop \cite{KorthalsAltes:1999xb,KorthalsAltes:2000gs}
is equal to the order-order interface tension for $Z(N)$ interfaces
\cite{Bhattacharya:1990hk, *Bhattacharya:1992qb}.
(Throughout this paper we assume that the 't Hooft loop is purely spatial;
with or without quarks,
temporal 't Hooft loops do not develop an area law at any
temperature \cite{deForcrand:2001nd}.)

Dynamical quarks carry $Z(N)$ electric charge, and so modify the behavior
of both loops.  The Polyakov loop is no longer a strict order parameter,
but is nonzero at any temperature $T \neq 0$.  Similarly, one expects that
the 't Hooft loop acts as in
the deconfined phase of the pure glue theory, and so exhibits an
area law.
However, it is not clear how to define
the 't Hooft loop in a theory with dynamical quarks 
\cite{KorthalsAltes:1999xb,KorthalsAltes:2000gs}.

In this paper we show that 't Hooft loops are well defined for a
Roberge-Weiss transition at $\phi_{RW} = k/(2 N)$.
Establishing this result is not difficult.  In a gauge
theory, a global $Z(N)$ transformation is an overall rotation of the quark
field by a constant phase.  This can be exactly compensated by a 
shift in an imaginary chemical potential for the quarks.  
What is less obvious is how the boundary conditions of the 
Roberge-Weiss transition
are precisely equivalent to those for a $Z(N)$
interface, and thus to the 't Hooft loop 
\cite{KorthalsAltes:1999xb,KorthalsAltes:2000gs}.

We compute the behavior of the 't Hooft loop at asymptotically high 
temperature, where the calculation can easily be done using semi-classical
techniques.  In the pure glue theory, at leading order and up to
corrections $\sim g^3$, the interface tension associated
with 't Hooft loops of different charges satisfies Casimir scaling
\cite{Giovannangeli:2002uv, *Giovannangeli:2004sg}.
We find that even to leading order, dynamical quarks do not respect
the Casimir scaling found in the pure glue theory.

We also consider the thermodynamics of Roberge-Weiss phase transitions.
We show that the interaction measure is negative 
on at least one side of the Roberge-Weiss transition. This happens
because both the chemical potential and the associated quark number
densities are imaginary.  Thus their contribution to the energy density
can be negative, and so unphysical.  

It is also of interest to know how the 't Hooft loop behaves for
temperatures which are not asymptotic.  To study this, we consider
heavy quarks, so that there are 
Roberge-Weiss and deconfining phase transitions
in the plane of temperature and imaginary
chemical potential.  We use a matrix model 
\cite{Dumitru:2011, *Dumitru:2012fw,Pisarski:2012bj}, 
which was used previously 
to locate the position of the deconfining critical endpoint
for heavy quarks \cite{Kashiwa:2012va}.
In this matrix model we find intersecting lines of
first order transitions in the $T - \phi$ plane.
There are lines of first order transitions both at $\phi_{RW}$
and $\phi \neq \phi_{RW}$.  Depending upon the quark mass,
these first order lines end in either critical or tri-critical points.
Our results are in good agreement with 
recent results on the lattice \cite{Fromm:2012yg}.

The phase transitions in $T$ and $\phi$ are naturally characterized by
a modified Polyakov loop, which is constructed to be invariant under
the Roberge-Weiss symmetry.  For
the transition at $\phi_{RW}$, as occurs at high temperature, only the
imaginary part of the Polyakov loop is discontinuous.  At lower temperatures,
there are first order transitions for $\phi \neq \phi_{RW}$, where
both the real and imaginary parts of the modified Polyakov loop jump.
Since the real part of the Polyakov loop characterizes the usual
phase transition for deconfinement, the
transitions when $0< \phi \neq \phi_{RW}< 1/N$ are those where
deconfinement mixes with the Roberge-Weiss transition.
We find that the interface tension of a first
order transition is related to a 't Hooft loop if and only if
it is entirely Roberge-Weiss, occurring at $\phi_{RW}$.  

In this paper we do not analyze a matrix model with light quarks,
where the restoration of chiral
symmetry also enters
\cite{Alford:1998sd, *Hart:2000ef, *deForcrand:2002ci, *D'Elia:2002gd, *D'Elia:2004at, *Chen:2004tb, *deForcrand:2006pv, *Wu:2006su, *D'Elia:2007ke, *deForcrand:2008vr, *D'Elia:2009tm, *D'Elia:2009qz, *deForcrand:2010he, *Cea:2012ev, Bonati:2011a, *Bonati:2011b, *Nagata:2011yf, *Wu:2013bfa}.
At high temperature there is only a Roberge-Weiss transition at $\phi_{RW}$,
while at lower temperatures, the order parameter for chiral symmetry
breaking can mix with the imaginary part of the Polyakov loop.
Nevertheless, we suggest that an analogous 
criterion occurs: the 't Hooft loop is related to the interface tension
of a first order transition only at high temperature, for a 
Roberge-Weiss transition at $\phi = \phi_{RW}$.

Hence in general, 't Hooft loops can only be defined 
at temperatures which are above that for deconfinement
or the restoration of chiral symmetry.
Nevertheless, we find it noteworthy that 
even a limited region of an extended phase diagram, it is
possible to measure 't Hooft loops in a gauge theory with dynamical quarks.

In Sec. (\ref{sec:rw_hooft}) we discuss the global $Z(N)$ symmetries
of the pure gauge theory, the symmetries with dynamical quarks,
and the relationship to 't Hooft loops.
In Sec.  (\ref{sec:thermo}) we consider the thermodynamics of
Roberge-Weiss phase transitions, and the phase diagram with heavy
quarks.

\section{$Z(N)$ symmetry and 't Hooft loops}
\label{sec:rw_hooft}
\subsection{General analysis}
\label{sec:general}

We work in imaginary time $\tau$ at a temperature $T$,
so $\tau: 0 \rightarrow 1/T$.  Under a gauge transformation $U(\vec{x},\tau)$,
the gluons and quarks transform as
\begin{equation}
A_\mu(\vec{x},\tau) \rightarrow 
\frac{1}{ig} \, U(\vec{x},\tau)\,  D_\mu  \, U^\dagger(\vec{x},\tau) \;\; ,
\;\;
q(\vec{x},\tau) \rightarrow U(\vec{x},\tau) \, q(\vec{x},\tau) \; ,
\end{equation}
$D_\mu = \partial_\mu + i g A_\mu$.

A special class of gauge transformations are those which are 
symmetric up to elements of $Z(N)$.  Consider the matrix
${\rm e}^{ 2 \pi i k/N} {\bold 1}_{N}$; this has determinant
one, and so is an element of $SU(N)$.  Since it is proportional to
the unity matrix, we can consider aperiodic gauge transitions
\begin{equation}
U(\vec{x},1/T) = {\rm e}^{ 2 \pi i k/N} U(\vec{x},0) \; ,
\label{zn_transf}
\end{equation}
where $k$ is an integer.  Gluons must be
periodic in $\tau$, but this aperiodicity in the $U$'s does
alter the gluon boundary conditions, cancelling between $U$ and $U^\dagger$.

Quarks are fermions, and so in the absence of a chemical
potential, are anti-periodic in $\tau$.
The boundary conditions are altered by
the presence of a chemical potential, $\mu$.
Consider the analytic continuation from a real chemical potential, 
$\mu$, to one which is imaginary and proportional to the temperature,
\beq
\mu = 2 \pi i \, \phi \, T \; .
\eeq
The quarks now satisfy
\beq
q(\vec{x}, 1/T) = - \, {\rm e}^{ - 2 \pi i \, \phi} \, q(\vec{x}, 0) \; .
\label{quark_bc}
\eeq
Because of the change in the fermion boundary conditions, when
$\phi \neq 0$ the theory can exhibit unphysical behavior.  
For example, $\phi = \frac{1}{2}$ turns fermions into bosons.  We shall
see an example of this unphysical behavior later, 
when we find that the interaction measure is negative
on one side of the Roberge-Weiss transition line.

As the $Z(N)$ transformation of Eq. (\ref{zn_transf})
is proportional to the unit matrix,
for the quarks it just alters them by 
an overall phase.  Thus in all, quarks transform as
\begin{equation}
q(\vec{x}, 1/T) = \, 
- {\rm e}^{ 2 \pi i \, (- \phi + k/N)} \, q(\vec{x}, 0) \; .
\label{q_bcs}
\end{equation}
It is useful to think of the boundary condition on the quarks at $\tau = 1/T$
as a type of $Z(N)$ ``charge''.  
Because the $Z(N)$ transformations of the gauge field arise from
the details of the $SU(N)$ algebra, this 
$Z(N)$ quark charge is heuristic, meant mainly to understand
the detailed computations in the next subsection.
By Eq. (\ref{q_bcs}) 
we normalize the $Z(N)$ charge of the quark as $- \phi + k/N$.

When there is an imaginary chemical potential, then, a shift
in $\phi$ can be compensated by a $Z(N)$
transformation,
\begin{equation}
\phi \rightarrow \phi + \frac{k }{N} \; .
\label{partition_sym}
\end{equation}
The action is invariant under this transformation.  By construction,
so are the boundary conditions.  
This is the Roberge-Weiss symmetry \cite{Roberge:1986}.

This periodicity can also be 
understood topologically.  In the pure glue theory,
the gauge group is $SU(N)/Z(N)$, and so there is a global symmetry of
$Z(N)$.  When $\phi = 0$,
the gauge group is just $SU(N)$, and this global $Z(N)$ symmetry is lost.
However, by 
introducing an imaginary chemical potential, $\phi$, for free quarks
we gain an extra global symmetry of $U(1)$.  The coupling of
gauge fields reduces this $U(1)$ symmetry to 
one of $Z(N)$.  

We thus have global $Z(N)$ symmetries both in the pure glue theory
and, when $\phi \neq 0$, for dynamical quarks.  Thus we might expect 
that there are non-contractible loops which measure windings in 
these global $Z(N)$ symmetries.
In the pure glue theory, such windings are measured by the 't Hooft loop.
Since the transformation of $\phi$ is so intimately tied to the
$Z(N)$ symmetry of the pure glue theory, it is perhaps not so surprising
that the 't Hooft loop continues to measure such windings, even in
the presence of dynamical quarks.  

We can understand where the Roberge-Weiss transition occurs even without
detailed computation.  By the Roberge-Weiss symmetry, $\phi = 0$ is
equivalent to $\phi = k/N$.  
The simplest way is to introduce a background $Z(N)$ charge
$k/N$, so that at $\phi = k/N$, the total $Z(N)$ charge is zero.
The question is then how to match these two cases, moving up from
$\phi = 0$, and down from $\phi = k/N$.  Since at some point the
background field has to jump from zero to $k/N$, 
this might generate a transition at some $\phi$.

The most natural place for the transition to occur
is exactly halfway in between, at
\begin{equation}
\phi_{RW} = \frac{k}{2 \, N} \; .
\label{RW_point}
\end{equation}
As $\phi$ increases from zero, the $Z(N)$ quark charge
decreases, going from zero at $\phi = 0$ to
$- \, \phi_{RW}$ just to the left of $\phi_{RW}$.

Now work in the opposite direction, letting $\phi$ decrease 
from $k/N$.  Here we have to include the background charge $k/N$,
so that as $\phi = k/N$, the total $Z(N)$ charge vanishes, consistent
with the Roberge-Weiss symmetry.  As $\phi$ decreases, the $Z(N)$ charge
increases, and equals
$-\, \phi_{RW} + k/N = + \, \phi_{RW}$ just to the right of $\phi_{RW}$.

Thus exactly at the Roberge-Weiss point, the quark $Z(N)$ charge is
$-\, \phi_{RW}$ on the left, and $+ \, \phi_{RW}$ on the right.  
Because of charge conjugation
symmetry, various quantities transform simply as we flip the sign
of the quark $Z(N)$ charge.  The pressure is even in $\phi$, so
quarks with equal and opposite $Z(N)$ charge are degenerate,
and have no forces acting between them.  
In contrast, the (imaginary) quark number density is odd in $\phi$,
and flips moving across $\phi_{RW}$.  The change in the sign
of the quark number density implies that if we change $\phi$ at 
fixed temperature, that there is a phase transition 
at $\phi_{RW}$ which is of first order.

The existence of a first order transition at $\phi_{RW}$ is valid only
at high temperature.  As we show in the next section using a matrix model
for heavy quarks, and in agreement with lattice
results \cite{Fromm:2012yg}, at temperatures 
near that for deconfinement there can be
transitions for $\phi \neq \phi_{RW}$, which are
of either first or second order.
A similar statement can
be made for light quarks, near the transition for the restoration of
chiral symmetry
\cite{Bonati:2011a, *Bonati:2011b, *Nagata:2011yf, *Wu:2013bfa}.

\subsection{Semi-classical analysis}
\label{sec:semi}

Although the above discussion of quark $Z(N)$ charge
is illustrative, it is merely heuristic.
This is because while the quark chemical potential only involves
a $U(1)$ phase, the $Z(N)$ transformations of the gauge field involve
the Lie algebra of $SU(N)$, and in particular
the Cartan subalgebra of mutually commuting generators, 
in a detailed manner.
In this subsection we perform an analysis at high temperature,
where all computations can be done semiclassically.  

To generate a $Z(N)$ transformation, consider the diagonal matrix
\cite{Giovannangeli:2002uv, *Giovannangeli:2004sg}
\beq
\bdy_k = \frac{1}{N}\; {\rm diag}
\left( k \ldots k, -N + k,
\ldots -N + k \right) \; .
\eeq
There are $N-k$ elements with entry $k/N$, and $k$ 
elements equal to $-1 + k/N$. 
Thus $Y_k$ has zero trace and is an element of the Lie algebra of $SU(N)$.  
Since
\beq
{\rm e}^{2 \pi i \bdy_k} = {\rm e}^{2 \pi i k/N} \, {\bold 1}_N \; ,
\eeq
$\bdy_k$ generates a $Z(N)$ transformation with strength $k$.
$\bdy_3$ is the hypercharge matrix for three flavors. 

The thermal Wilson line is given by
\beq
{\bf L}(\vec{x}) = \exp\left( i g \int^{1/T}_0 \, A_0(\vec{x},\tau) 
d \tau \right) \; .
\label{wilson_line}
\eeq
Then the path 
\begin{equation}
A_0 = \frac{2 \pi T}{g} \; q \; \bdy_k \; ,
\end{equation}
takes one from the ordinary perturbative vacuum for $q = 0$,
${\bf L} = {\bf 1}_N$, to the $k$th 
$Z(N)$ transform thereof,
for $q = 1$, ${\bf L} = {\rm e}^{2 \pi i k/N} {\bf 1}_N$.

At the outset it is also useful to consider 
the matrix $\bdy_{N - k}$, where
\beq
\bdy_{N - k} = \frac{1}{N} {\rm diag}
\left( N - k \ldots N - k, - k \ldots - k \right) \; ;
\eeq
this has $k$ elements with entry $1-k/N$, and $N-k$ elements with
value $-k/N$.  
%
%
%
It is clear that
\beq
{\rm e}^{2 \pi i \bdy_{N - k}} = 
{\rm e}^{2 \pi i (- k)/N} \, {\bold 1}_N \; .
\eeq
That is, $\bdy_{N-k}$ generates a $Z(N)$ transformation
with charge $N-k$, which by the additive $Z(N)$ symmetry is equivalent to
charge $-k$.  
This symmetry is useful, because we expect that the interface tension for
charge $k$ should be equal to that for charge $N-k$.  
The matrix $\bdy_{N-k}$ makes this manifest, 
although our final expressions do not obviously
reflect this symmetry.  Because of this symmetry, we can restrict
$k$ to be less than the nearest integer $\leq N/2$.  

Classically there is no potential for the $q$'s, but one is generated
at one loop order.  This was first computed 
by Weiss \cite{Weiss:1981} and Yaffe {\it et al.} \cite{Gross:1981br},
\beq
{\cal V}_\mathrm{pt}^{gl}
= \frac{2 \pi^2 T^4}{3} \left(
- \; \frac{N^2 - 1}{30}
+ \sum_{i,j=1}^{N} 
   q_{ij}^2 (1-|q_{ij}|)^2 \right) \; .
\label{pert_potential}
\eeq
The $q_i$ are the elements of $q \, \bdy_k$.
The gluon contribution involves the adjoint covariant derivative, and so
only the differences of the $q_i$ enter, $q_{i j} = q_i - q_j$.

The $q_{i j}$'s arise as the gluon energies divided by $2 \pi T$.
Since the potential for the $q_i$'s arises for a Matsubara sum, then
it is periodic in the $q_{i j}$, $q_{i j} \rightarrow q_{i j} + 1$.  
Consequently, the potential 
only involves the absolute value of $q_{i j}$, modulo one.  Thus for
the gluon term we can always require that $q: 0 \rightarrow 1$.

The term independent of $q$
is the just (minus) the pressure of an ideal gas of gluons.  
For the $q_{i j}$, there are $k$ elements $=q(k/N)$, 
and $N-k$ elements $= q(-1 + k/N)$.  The potential arises from
the (absolute value) of the $q_i - q_j$'s.  For these elements,
the $|q_{i j}|$'s are either zero, or $q$ (assuming $q > 0$).  
There are $2 k (N-k)$ such terms which give $q$, and 
so the gluon potential is
\beq
{\cal V}_\mathrm{pt}^{gl}(q) - {\cal V}_\mathrm{pt}^{gl}(0)
= \frac{4 \pi^2 T^4}{3} \; k(N-k) \; q^2 (1 - q)^2 \; .
\label{gluon_potential}
\eeq
The potential is identical at $q = 0$ and $q = 1$, which
illustrates
the $Z(N)$ symmetry of the pure glue theory between
the ordinary vacuum, $k = 0$, and the $k$th $Z(N)$ transform.  

In all, the potential is proportional to $k (N-k)$.  This is known as
Casimir scaling.  It is satisfied in the pure glue theory up to
corrections $\sim g^3$ times the leading order term
\cite{Giovannangeli:2002uv, *Giovannangeli:2004sg}.

For a single, massless flavor,
quarks contribute to the potential for $q$ as
%
%
\beq
{\cal V}_\mathrm{pt}^{qk}
= -\frac{4 \pi^2 T^4}{3} \left(
- \; \frac{N}{30}
+ \sum_{i=1}^{N} 
   q_{i}^2 (1-|q_{i}|)^2 \right) \; .
\label{qk_potential}
\eeq
For the $q_i$, there are $N-k$ elements 
$$
q_i = \frac{1}{2} - \phi + \frac{k}{N} \, q  
\; , 
$$
and $k$ elements
\beq
q_i = \frac{1}{2} - \phi + \left(-1 + \frac{k}{N}\right) q \; .
\; ,
\eeq
These $q$'s differ from the gluon case because of the change
in the boundary conditions in $\tau$, for fermions and bosons, and
because the quarks carry an imaginary chemical potential $\sim \phi$.
The quark contribution involves the covariant derivative in the
fundamental representation, so it is the $q_i$,
and not $q_i - q_j$, which enter.  Like
the gluon potential, 
as the quark potential arises from a sum over the Matsubara frequencies,
each $q_i$ enters only as the absolute value, modulo one.  
Thus it is necessary to be careful about the range of the $q_i$.
Even so, clearly Eq. (\ref{qk_potential})
is invariant under two symmetries.  As only the absolute value
of the $q_i$ enter, one is $q_i \rightarrow - q_i$.
The second is apparent from the form of the potential: assuming
that $1 \geq q_i \geq 0$, the quark potential is also invariant under
$q_i \rightarrow 1 - q_i$.

The $q_i$ are the Euclidean energies divided by $2 \pi T$.  
Thus the first factor of $\frac{1}{2}$ is the $\pi T$
which arises because quarks are fermions, with boundary
conditions which are anti-periodic in imaginary time.
The second term,
$- \phi$, is the contribution of the imaginary chemical potential.
Lastly, the terms $\sim q$ arise from the background field.

While we explicitly compute the potential shortly, most aspects of the physics
can be understood without going into such details.

First consider vanishing chemical potential,
$\phi = 0$.  As discussed above, for the gluons the potential is
periodic in $q$, with $q=1$ degenerate with $q=0$, Eq. (\ref{gluon_potential}).
This reflects the $Z(N)$ symmetry of the pure glue theory.

This is no longer true with dynamical quarks.  Then $q_i$'s involve
factors of $kq/N$; for $q = 1$, this is $k/N$,
and generates a nontrivial 
potential.  Computation shows the quark potential
has a higher value when $q = 1$ than for $q = 0$.  This occurs
because quarks in the fundamental representation do not respect the
$Z(N)$ global symmetry of the pure glue theory.

Looking just at the $q_i$'s shows how the Roberge-Weiss
symmetry of Eq. (\ref{partition_sym}) 
works.  Compare the ordinary perturbative vacuum, where
$\phi = q = 0$, to a state where $\phi = k/N$, and $q = 1$.
When $\phi = q = 0$, all $q_i$'s equal $\frac{1}{2}$.
When $\phi = k/N$, and $q = 1$, 
all of the $q_i$'s are either $\frac{1}{2}$ or $- \frac{1}{2}$.
Since only the absolute value enters, both are 
equivalent to $\frac{1}{2}$.  

Thus we see that a state with $\phi = k/N$ is equivalent to the
perturbative vacuum, if we shift the background field by $k/N$.
As we increase $\phi$ from $0$, there is no background
field, but at some point, it shifts to that with $k/N$.  As discussed
previously, this happens halfway in between, at $\phi_{RW} = k/(2N)$.  

At $\phi_{RW}$, when $q = 0$ all $N$ elements equal
\beq
q_i = \frac{1}{2} \left( 1 - \frac{k}{N} \right)  \; .
\label{RW_left}
\eeq
When $q = 1$, there are $N-k$ elements equal to
\beq
q_i = \frac{1}{2} \left( 1 + \frac{k}{N} \right)  \; ,
\label{RW_rightA}
\eeq
and $k$ elements equal to 
\beq
q_i = - \frac{1}{2} \left( 1 - \frac{k}{N} \right)  \; .
\label{RW_rightB}
\eeq
Even without computation one can see that the values
in Eqs. (\ref{RW_rightA}) and (\ref{RW_rightB}) are equal to those in
in Eq. (\ref{RW_left}).
The $q_i$'s in Eq. (\ref{RW_rightA}) are related to those in
in Eq. (\ref{RW_left}) by $q_i \rightarrow 1 - q_i$; 
the $q_i$'s in Eq. (\ref{RW_rightB}) are equivalent to
those in Eq. (\ref{RW_left}) under $q_i \rightarrow - q_i$.  

Explicitly, at $\phi_{RW} = k/(2 N)$ the quark potential is given
by
$$
{\cal V}_\mathrm{pt}^{qk}(q) = - \frac{4 \pi^2}{3} \, T^4 \, 
\left( ( N - k) 
\left( \frac{1}{4} - \left( \frac{k}{N}\right)^2 
\left( \frac{1}{2} -  q \right)^2  \right) 
\right.
$$
\beq
\left.
+ \; k \left( 1 - \frac{k}{N} \right)^2 \left(\frac{1}{2} -q \right)^2
\left( 1 - \left(1 - 
\frac{k}{N} \right) \left|\frac{1}{2} - q\right| \right)^2 \right) \; .
\label{quark_potential}
\eeq
The values of the potential are clearly equal when $q = 0$ and $q = 1$;
note, however, that the second term does involve the absolute value
of $\frac{1}{2} - q$.  
Also, this potential is not simply proportional to $k (N-k)$, and so
does not respect Casimir scaling.  As argued above, the potential
will respect a transformation under $k \rightarrow N-k$, since then
$\bdy_{-k}$ enters, instead of $\bdy_k$.

We can use these results to discuss 't Hooft loops.
In the theory without dynamical quarks, Kovner, Korthals-Altes, and
Stephanov \cite{KorthalsAltes:1999xb}
showed that the 't Hooft loop has a simple physical interpretation.
Consider a box which is long in one spatial direction, say that in the
$z$-direction.  Put a 't Hooft loop of $k$th $Z(N)$ charge
around the boundary of the box at one end of the box, at 
$z = L$.  
Then consider boundary conditions which are $q=0$ at $z = 0$, and
$q=1$ at $z=L$.  These boundary conditions
are identical to that for a order-order $Z(N)$ interface of charge $k$
\cite{Bhattacharya:1990hk, *Bhattacharya:1992qb, Giovannangeli:2002uv, *Giovannangeli:2004sg}.  
The 't Hooft loop at $z=L$ then forces $q$ to jump from $1$ back to $0$,
so that in all one has periodic boundary conditions.  Neglecting this
singularity shows that the interface tension for the order-order
interface is equal to that for the 't Hooft loop. 

Our analysis above shows that 
one has identically the same boundary
conditions across the Roberge-Weiss transition point.
The vacuum jumps from $q = 0$ on the
left hand side, to $q = 1$ on the right hand side.  Consequently,
at $\phi_{RW}$, we can define the 't Hooft loop
in precisely the same manner as in the pure gauge theory.

We stress that the 't Hooft loop can {\it only} be defined 
at the Roberge-Weiss transition point, when $\phi_{RW}$.
For example, in the ordinary vacuum, $\phi = 0$, 
the states with $q = 0$ and $q = 1$ are not degenerate.
Typically, the state with $q = 1$ is not even
extremal \cite{KorthalsAltes:1999xb,KorthalsAltes:2000gs}.  (Depending
upon the matter content, it is possible that the state with $q = 1$
is metastable \cite{Belyaev:1991np}; see, also,
\cite{Smilga:1993vb}.)
In this case, one can define a 't Hooft loop, but the
interface tension will have an imaginary part, reflecting this metastability.
A necessary condition for the $Z(N)$ loop to be non-contractible is
if the $Z(N)$ transformed states are absolutely degenerate.
This only happens across a Roberge-Weiss transition.

In the pure gauge theory, the value of the 't Hooft loop depends only
upon the area of the loop, but not upon its shape.  Away from the 
Roberge-Weiss transition, the $Z(N)$ charges of the quarks are unequal,
and have different pressures.  This difference in pressure generates a
force, which drives the 't Hooft loop to be flat.

In contrast, at the Roberge-Weiss transition the $Z(N)$ charges of the
quarks are equal and opposite, and have the same pressure.
Thus there is no net force which acts upon the 't Hooft loop,
and its value depends only upon the area of
the loop, and not its shape.

In the pure gauge theory, the 't Hooft loop measures a 
non-contractible loop for the global $Z(N)$ symmetry.  As we argued
in Sec. (\ref{sec:general}), there is also a global $Z(N)$ symmetry
if the quarks have an imaginary chemical potential.  Since this
global $Z(N)$ symmetry of dynamical quarks is intimately tied to
the $Z(N)$ transformations of the pure glue theory, it is natural that
the 't Hooft loop continues to measure the winding in $Z(N)$.

Our results are valid at asymptotically high temperature, where the
Roberge-Weiss phase transition is manifestly of first order.  As one
lowers the temperature, numerical simulations on the lattice
show that the single transition at $\phi = \phi_{RW}$ can split into two,
for $\phi \neq \phi_{RW}$, which are of either first or second
order 
\cite{Alford:1998sd, *Hart:2000ef, *deForcrand:2002ci, *D'Elia:2002gd, *D'Elia:2004at, *Chen:2004tb, *deForcrand:2006pv, *Wu:2006su, *D'Elia:2007ke, *deForcrand:2008vr, *D'Elia:2009tm, *D'Elia:2009qz, *deForcrand:2010he, *Cea:2012ev, Bonati:2011a, *Bonati:2011b, *Nagata:2011yf, *Wu:2013bfa, Fromm:2012yg}.

In the next section we also find that for heavy quarks in a matrix model,
a single line of Roberge-Weiss transitions at $\phi_{RW}$
splits into two transitions for $\phi \neq \phi_{RW}$, see
Fig. (2).  This agrees with the lattice results of
Ref. \cite{Fromm:2012yg}.  We argue there that the 
interface tension across a first order transition is related to the
't Hooft loop if and only if the transition is entirely
Roberge-Weiss.  For a transition at $\phi_{RW}$,
the two degenerate
vacua are $Z(N)$ transformations of one another, and so
the interface tension is naturally related to the 't Hooft loop.
In contrast, for transitions for $0 < \phi \neq \phi_{RW} < k/N$,
the Roberge-Weiss and deconfining transitions mix with one another.
Thus across the dotted lines in Fig. (2), when $\phi \neq \phi_{RW}$
while the vacua are degenerate across the
transition, they are not $Z(N)$ transformations of one another.  
Thus the corresponding interface tension is not related to a 't Hooft loop.

For light quarks, numerical simulations on the lattice suggest that the
single line of Roberge-Weiss transitions for $\phi_{RW}$ at high temperature
can also
split into two transitions at low temperature, with $\phi \neq \phi_{RW}$
\cite{Bonati:2011a, *Bonati:2011b, *Nagata:2011yf, *Wu:2013bfa}.
When this happens, the Roberge-Weiss transition mixes with that for chiral
symmetry restoration.  We suggest that as for heavy quarks,
the interface tension across such a first order transition is related to
a 't Hooft loop only when $\phi = \phi_{RW}$.

\subsection{Computing the 't Hooft loop}

The explicit computation of the 't Hooft loop follows standard methods.
A $Z(N)$ interface has an electric field, which contributes to the
action as
\beq
\frac{4 \pi^2 T^2}{g^2 N} \; k (N- k) \; \int dz \; 
\left( \frac{d q}{dz} \right)^2 \; .
\label{kinetic}
\eeq
The interface tension is determined as a semiclassical tunneling between
$q = 0$ and $q = 1$.  In the pure glue theory, this involves the sum
of the kinetic term in Eq. (\ref{kinetic}) and the potential of
Eq. (\ref{gluon_potential}).  Since each term is proportional to
$k (N-k)$, the interface tension is as well, and so it respects
Casimir scaling, at least to leading order in the coupling constant.

With dynamical quarks, the kinetic term remains as in Eq. (\ref{kinetic}),
but now the potential is a sum of the gluon term in 
Eq. (\ref{gluon_potential}) and the quark term in Eq. (\ref{quark_potential}).
Since the quark potential is not proportional to $k (N-k)$, the interface
tension is a more complicated function of $k$.  

The appearance of Casimir scaling
is to some extent an observation about the structure of the theory in
weak coupling.  Computation to corrections $\sim g^3$ beyond that of
leading order shows a small violation of Casimir scaling even in the pure
glue theory \cite{Giovannangeli:2002uv, *Giovannangeli:2004sg}.  
So there is nothing fundamental in Casimir scaling, nor
in that it is violated by dynamical quarks.
The only symmetry principle which must be respected is that
for $Z(N)$ periodicity.  This requires that the interface tension
for $k$ is the same for $N-k$; this is equivalent to $-k$, which as we
have argued, is valid.

Let us assume that the path
for the interface tension is along the direction ${\bf Y}_k$.
This is a straight line in the Cartan subalgebra.
If true, it is easy computing the 't Hooft loop.
One has a 
tunneling problem in this one dimension, and it is easy to
solve this by using ``energy'' conservation for the
associated problem in quantum mechanics \cite{Bhattacharya:1992qb}.  
This involves both the potential and the kinetic term for the gluons,
Eq. (\ref{kinetic}).  For the pure glue theory, as both terms 
are $\sim k(N-k)$, and the interface tension follows immediately.
With dynamical quarks, since the quark contribution to the potential
is more involved, and even with the conservation of energy
one is left with a single integral over $q$ which needs to
be computed numerically.  We defer this exercise, and simply observe
that with dynamical quarks, Casimir scaling will not be satisfied.

This assumes that the straight line path is minimal.  We consider
this in the next subsection.

\subsubsection{Straight line path}

While in the pure glue theory the path
for the interface tension is along the direction ${\bf Y}_k$,
in principle it can move in other directions.  
The general path for an $SU(N)$ gauge theory lies
in the subspace of all commuting generators, which
is the Cartan subalgebra, with $N-1$ dimensions.

For the pure gauge theory one can show that the straight line path
is minimal
\cite{Giovannangeli:2002uv, *Giovannangeli:2004sg}.  
This has deep geometric reasons: 
the ${\bf Y}_k$ form the boundary
of the Weyl chamber, which is the smallest possible region to
describe the Cartan subalgebra \cite{Dumitru:2012fw}.
The endpoints of ${\bf Y}_k$ are the relevant endpoints for the interface
tension, so then 
it is natural that the boundary, along ${\bf Y}_k$, is the minimal path
which connects these two points.

With dynamical quarks, the structure of the Weyl chamber 
at the Roberge-Weiss transition is more involved.  
We have not been able to answer this question for an arbitrary
numbers of colors, and so
satisfy ourselves with working out the simplest possible cases,
working up from $N=2$ to $N=4$.

The case of two colors is trivial.  There is
only one direction, along ${\bold Y}_1 \sim \sigma_3 \sim {\rm diag}(1,-1)$,
and so the path necessarily lies along ${\bold Y}_1$.

The first nontrivial case arises for three colors, where there are two
directions in the Cartan sub-algebra.  For the quarks, the $q_i$'s are
\beq
q_i = \left(\frac{1}{2} - \phi\right) {\bold 1}_3
 + \frac{q_3}{2} \; {\rm diag}(1,-1,0)
 + \frac{q_8}{3} \; {\rm diag}(1,1,-2) \; .
\label{three_colors_path}
\eeq
In the standard Gell-Mann notation, the directions are
$\lambda_3 \sim (1,-1,0)$ and $\lambda_8 \sim {\bold Y}_1 \sim (1,1,-2)$, 
with associated coordinates $q_3$ and $q_8$.  For three colors,
there is only one interface tension, as $k = -2$ is equivalent
to $k = 1$.  The endpoints of the interface 
are given by $q_8 = 0$ and $1$, with $q_3 = 0$.  
The straight line path is along ${\bold Y}_1$, with one transverse
direction, along $\lambda_3$.

The $q_i$'s of Eq. (\ref{three_colors_path}) are for quarks, but we
can use them for gluons, since only the differences of the
$q_i$'s enter in the gluon potential, through $q_{i j} = q_i - q_j$.  
Numerically we find that for 
the gluon potential of Eq. (\ref{pert_potential}),
the straight line path along ${\bf Y}_1$ is minimal.
That is, for the path where $q_8 \neq 0$ and $q_3 = 0$, 
for every value of $q_8$
the potential is minimal with respect to variations in the transverse
direction, along $q_3$.
As noted, this is because ${\bf Y}_1$ is the boundary of the Weyl chamber
for three colors.

Now consider the quark potential at the Roberge-Weiss
transition point, $\phi = 1/6$, using the $q_i$'s of Eq.
(\ref{three_colors_path}) in Eq. (\ref{qk_potential}).
As in the pure glue theory for three colors, numerically we have checked
that the minimal path is a straight line along ${\bold Y}_1$.  

For four colors there are two possible interfaces, 
$k=1$ and $k=2$; the associated elements of $Z(4)$ are
$i$ and $-1$, respectively.  

For $k = 1$ we can parametrize the $q_i$'s 
using the usual Cartan generators,
\beq
q_i = \left(\frac{1}{2} - \phi\right) {\bold 1}_4
 + \frac{q_3}{2} \; {\rm diag}(1,-1,0,0)
 + \frac{q_8}{3} \; {\rm diag}(1,1,-2,0) 
+ \frac{q_{15}}{4} \; {\rm diag}(1,1,1,-3)  \; .
\label{four_colors_pathA}
\eeq
The $k=1$ interface is from $q_{15}: 0$ to $1$, with
$q_3 = q_8 = 0$ at either end.  The 
straight line path is along ${\bold Y}_1 \sim (1,1,1,-3)$, with
only $q_{15}$ nonzero.  
Numerically we checked that the straight line path
is minimal, both in the pure glue theory and with dynamical quarks
at the Roberge-Weiss point for $k=1$, where $\phi_{RW} = 1/8$.  

This exercise also shows that the potential has nontrivial structure.
At the Roberge-Weiss transition for $k=1$, $\phi_{RW} = 1/8$,
there is a metastable minimum in the potential when
$q_{15} = 0$: it occurs for $q_8 = 1$, with
$q_3 = 0$.  It is metastable in all three directions, but tunnels 
with finite lifetime to
the usual $Z(4)$ vacua, which are absolutely stable.  Such metastable
vacua are known to arise for these types of potentials
\cite{Belyaev:1991np}.

Lastly we consider four colors with $k = 2$.  For the $q_i$'s we take
\beq
q_i = \left(\frac{1}{2} - \phi\right) {\bold 1}_4
 + \frac{q_2}{2} \; {\rm diag}(1,-1,0,0)
 + \frac{q_2'}{2} \; {\rm diag}(0,0,1,-1) 
+ \frac{q_4}{4} \; {\rm diag}(1,1,-1,-1)  \; .
\label{four_colors_path_B}
\eeq
For the interface with $k=2$, $q_4 = 0$ at one end and $q_4=1$ at the other,
with $q_2 = q_2' = 0$ at both ends.  The straight line path is
along ${\bold Y}_2 \sim (1,1,-1,-1)$.  

To determine stability of a path it is essential to have
a parameterization in three independent directions.  
Our path is along ${\bold Y}_2 \sim (1,1,-1,-1)$.  The
diagonal matrix $\sim (1,-1,0,0)$ is a generator for 
(the diagonal part of) $SU(2)$ in the first
two colors, and is obviously transverse to ${\bold Y}_2$.
To determine the remaining direction, one can use 
brute force: one computes the 
linear combination of $(1,1,-2,0)$ and $(1,1,1,-3)$
which is transverse to $(1,-1,0,0)$ and
$(1,1,-1,-1)$. The answer, as in Eq. (\ref{four_colors_path_B}), is
$\sim (0,0,1,-1)$.  This is just
the $SU(2)$ type generator for the third and fourth 
and fourth colors of $SU(4)$.  At least after the fact, this is obvious.

By explicit computation, again 
one finds that the minimal path is a straight line, with
$q_4 \neq 0$ and $q_2 = q_2'=0$.  This is true
both for the pure glue theory, and for the theory with (massless)
dynamical quarks at the Roberge-Weiss transition point, $\phi_{RW} = 1/4$.
We did not find metastable minima when $k = 2$.

The examples of $k=1$ for three colors, and $k = 1$ and $k=2$ for
four colors, suggests that for $SU(N)$ at the Roberge-Weiss
transition point(s), $\phi_{RW} = k/(2N)$, that the path for the associated
interface tension is {\it always} a straight line along ${\bold Y}_k$.  
At present, we can only suggest this as a conjecture and have no
general proof.  If true, surely it is due to
the nature of the Weyl chamber with dynamical quarks at
the Roberge-Weiss transition point(s).

\section{Thermodynamics of Roberge-Weiss transitions}
\label{sec:thermo}
\subsection{High temperature}
\label{sec:high}

The total pressure $p$ and the entropy density $s$ are a sum,
\begin{align}
p = p_g + p_q,~~~~
s = s_g + s_q,
\end{align}
where $p_g$ and $s_g$ are the gluon contributions, and
$p_q$ and $s_q$ the quark contributions.

On the left side of the Roberge-Weiss transition, 
$\phi_{RW} = k/(2N)$,
the gluon contribution is
\begin{align}
p_g &=  
(N_\mathrm{c}^2-1) \frac{\pi^2 T^4}{45}\;\; ; \;\;
s_g  = 4 \; \frac{p_g}{T},
\end{align}
while the quarks contribute 
\begin{align}
p_q &=
\frac{\pi^2 N N_\mathrm{f}T^4 }{3}
\Bigl[ \frac{7}{60} -
2\Bigl(\frac{k}{N}q-\phi\Bigl)^2 + 4 \Bigl( \frac{k}{N}q-\phi \Bigr)^4
 \Bigr] \;\; ;
\nonumber\\
s_q &= 4 \; \frac{p_q}{T} \;\; ;
\nonumber\\
\mathrm{Im} (n_q) 
&= \frac{2\pi N N_\mathrm{f} T^3 }{3}
\Bigl[ \Bigl( \frac{k}{N}q-\phi \Bigr) - 4 \Bigl( \frac{k}{N}q-\phi
 \Bigr)^3 \Bigr],
\label{eos1}
\end{align} 
Here $kq/N-\phi$ is defined between $-1/2$ and $1/2$.
The quark number density is imaginary because the chemical potential is.

To define the energy density, we take the standard expression for
a real chemical potential, and assume it remains valid for
an imaginary chemical potential:
\beq
e = - p + sT + \mu n_q = - p + sT - 2\pi T \; \phi \; \mathrm{Im} (n_q).
\label{energy_density}
\eeq

Now consider how thermodynamic functions change on 
either side of the Roberge-Weiss transition, $\phi_{RW} = k/(2N)$.
On the left side of the transition, $q = 0$, while on the right,
$q = 1$, so $kq/N - \phi = \mp k/(2N)$ changes sign across the
transition.  The pressure and the entropy density are even
in $kq/N - \phi$ and do not change.
The quark number density is odd in $kq/N - \phi$
and so changes sign.  This change in sign for the (imaginary)
quark number density is the only reason why the energy density changes
at the Roberge-Weiss point and makes the transition of first order.

If we assume that $\phi$ and $T$ are both fixed,
at asymptotically high temperature the contribute of the quark
number density to the internal energy density is:
\begin{align}
\frac{e}{T^4}
&\to 3 \frac{p} {T^4} - 2\pi \phi \frac{\mathrm{Im} (n_q)}{T^3}
\label{ene}
\end{align}
Thus at high $T$ the interaction measure is due entirely to
the contribution from the quark number density,
\begin{align}
\Delta \equiv
\frac{e-3p}{T^4} &= - 2\pi \phi \frac{\mathrm{Im} (n_q)}{T^3}.
\label{IM}
\end{align}
The interaction measure is nonzero even at
high $T$ because we assume that the (imaginary) chemical
potential is proportional to temperature, $\mu = 2 \pi i T \phi$.
As discussed, it also flips sign across the transition.  This
holds not only at high $T$, but
persists down to temperatures close to the transition temperature,
as we see in the model calculations which follow.

\subsection{Non-perturbative models}
\label{sec:matrix}

The perturbative potential for the $q$'s is given by 
Eq. (\ref{pert_potential}).  It involves the function
$V_2(x) = x^2 (1 - |x|)^2$, and $q_{ij} = (q_i -
q_j)_{\mathrm{mod~1}}$. 

The minimum of this potential is always the usual perturbative vacuum,
or a $Z(N)$ transform thereof.  To model the transition to deconfinement,
we can add, by hand, non-perturbative terms
\cite{Dumitru:2011, *Dumitru:2012fw,Pisarski:2012bj,Kashiwa:2012va}.
The involves one new function, $V_1(x) = |x| (1-|x|)$, where like
$V_2(x)$, this function is defined to be periodic in $x$, modulo one.  

For all values of $N$, a successful fit was obtained with the form
\beq 
{\cal V}^{np}_g = - \frac{4 \pi^2}{3} \; T^2 \, T_d^2 
\sum_{i,j} \left( - \frac{c_1}{5} \; 
V_1(q_{ij}) - c_2 \; V_2 (q_{ij}) + 
\frac{N^2-1}{60} \; c_3 \right)
\label{non_pert_pot}
\eeq
The parameters $c_1$ and $c_2$ are assumed to be independent of temperature,
while the temperature dependence of $c_3$ is simply
\beq
c_3(T)
= c_3(\infty) + (c_3(T_d)-c_3(\infty))
\left(\frac{T_d}{T}\right)^2 \; .
\label{c3_def}
\eeq
Driven by the lattice data, most terms in the non-perturbative potential
are $\sim T^2$.  The temperature dependence of $c_3$ also incorporates
a MIT bag constant $\sim c_3(T_d)-c_3(\infty)$.  

The parameters of the model are fit by the transition in the pure glue
theory, where we assume $T_d = 270$~MeV.  At the outset, the model
involves four parameters, $c_1$, $c_2$, $c_3(T_d)$ and $c_3(\infty)$.
One then requires two conditions.  First, that the transition occurs
at $T_d$.  Second, that the pressure vanishes at $T_d$.  The first
is a reasonable condition at any $N$.  The second is an approximation,
modeling that the pressure is suppressed by $1/N^2$ in the confined phase,
relative to that in the deconfined phase.  
This leaves two parameters, which are fit by the value of the latent
heat, and the fall off of the pressure with temperature at
asymptotically high $T$.

To include quarks, we follow Ref. \cite{Kashiwa:2012va} and simply
add the one loop term for quarks in the background field.  
This is then the only way in which the (imaginary) chemical potential
enters.  For constant quark mass, this contribution is
\begin{align}
{\cal V}_q
& = -2 T \mathrm{tr}_{f,c} 
    \int \frac{dp^3}{(2\pi^3)} 
      \Bigl[
      \ln \Bigl( 1 + e^{-\beta \{ E_f - i 2 \pi T (\phi + q_c) \}}
          \Bigr)
    + \ln \Bigl( 1 + e^{-\beta \{ E_f + i 2 \pi T (\phi + q_c) \}}
          \Bigr)
    \Bigr],
\label{perq}
\end{align} 
where $\mathrm{tr_{f,c}}$ represents the trace over flavor and
color and $E_f(p) = \sqrt{p^2 + m_f^2}$. 
We assume that the up and down quarks are isospin symmetric.
We thus adopt the notation
that the light (up and the down) quark masses are $m_l$,
and the
strange quark mass is $m_s$. 

For light quark masses there is a back from
to the gluon potential for chiral symmetry breaking or restoration,
but we can neglect this here.
Such back reaction may be discussed by using the functional renormalization
group or the gluon and ghost potentials from the Landau gauge gluon and
ghost propagators~\cite{Fukushima:2012}. When we consider the small
quark mass region, we should also consider the meson and also the baryon
contributions, but inclusion of these effects is much more involved,
and will be treated later.

We shall work in the limit of heavy quark masses, where the light and strange
quark masses are constant, and such effects can be neglected.
Therefore, as a first step to construct the reliable model of QCD, we
investigate the upper part of the Columbia plot 
at the Roberge-Weiss endpoint.

From the thermal Wilson line $\bf L$ in Eq. \eqref{wilson_line}, we introduce
a modified Polyakov loop as
\begin{align} \Psi  &= {\rm e}^{2 \pi i\phi} \; 
\frac{1}{N} \; {\rm tr} \; {\bf L} \; ,~~~~ 
{\overline \Psi} = {\rm e}^{- 2 \pi i\phi} \;
\frac{1}{N} \; {\rm tr} \; {\bf L}^\dagger \; . 
\label{modified_Polyakov_loop}
\end{align} 
The Polyakov loop is the trace of the quark propagator.  Multiplying
by ${\rm e}^{2 \pi i \phi}$ is exactly like the phase which enters for
the boundary conditions for a dynamical quark field, Eq.
\eqref{quark_bc}.  The modified loop $\Psi$ is periodic
under the Roberge-Weiss symmetry, and so henceforth, we always refer
to the Polyakov loop as that of Eq. (\ref{modified_Polyakov_loop}).

Previously it was found the the
pressure and other thermodynamical quantities exhibit unphysical
behavior below $T_d$ in the matrix model \cite{Kashiwa:2012va}.  
This is because the
behavior of color singlet quantities, such as glueballs, are not
included self-consistently.  To handle this, we modify the potential
as
\beq
{\cal V}(q)  \rightarrow
{\cal V}(q) - {\cal V}(q_c) + {\cal V}(q_c) {\overline \Psi} \Psi \; ,
\label{modified_potential}
\eeq
where $q_c=(1/3,-1/3,0)$.
This follows a similar modification in Ref. \cite{Kashiwa:2012va}
where the $q$-dependent part of the potential was modified in one
particular direction in color space.  The above form is more natural,
and suppresses it in a color symmetric manner.

\subsection{Phase diagram for heavy quarks}
\label{sec:phase_heavy}

In this subsection we discuss the nature of the phase diagram, in
the plane of temperature and quark chemical potential, as the mass
of a heavy quark changes.  For this purpose, we can concentrate on the
behavior of the order parameter(s), which are the real and imaginary
parts of the (Roberge-Weiss symmetric) Polyakov loop.
After this we give results for thermodynamic quantities
in the following subsection.  
To be definite, we assume that there are
three degenerate flavors of quarks, although qualitatively
our results rather insensitive to the number of flavors.  

Consider first quarks in the
absence of an imaginary chemical potential, $\phi = 0$.  For
heavy quarks, we consider how deconfinement changes as we go from
the pure gauge theory, $m = \infty$, to large quark masses.  
For three (or more) colors, the deconfining transition is of first
order.  Quarks act like a background $Z(3)$ field for the transition,
and so in all parts of the phase diagram,
the real part of the Polyakov loop is nonzero whenever
there are dynamical quarks present.

As the quark mass decreases, the latent heat of the deconfining
transition decreases, until it first vanishes at 
a deconfining critical endpoint, when $m = m_{dce}$.  
Numerically, in the matrix model we find that
$m_{dce}/T_d \approx 8.2 \pm .1$ \cite{Kashiwa:2012va}. 
When $m < m_{dce}$, there is no deconfining transition, only crossover.
At the critical endpoint, $m = m_{dce}$.
the critical field is
the deviation of the real part of the Polyakov loop from its
expectation value, with the universality class that of the Ising model
in three dimensions.  In the matrix model
this occurs at a temperature which
is very close to that the transition temperature in the pure gauge theory,
$T_{dce}/T_d \approx .991 \pm .001$ \cite{Kashiwa:2012va}. 

When the quarks have an imaginary chemical potential which is nonzero, we
expect that the imaginary part of the Polyakov loop is also nonzero.  By
Roberge-Weiss periodicity, the imaginary part vanishes when $\phi = 0$
and $1/3$.  
In all cases, we find that the imaginary part of the Polyakov loop is
positive when $\phi < 1/6$, and negative when $\phi > 1/6$.  

\begin{figure}[htbp]
\begin{center}
\includegraphics[width=0.6\textwidth]{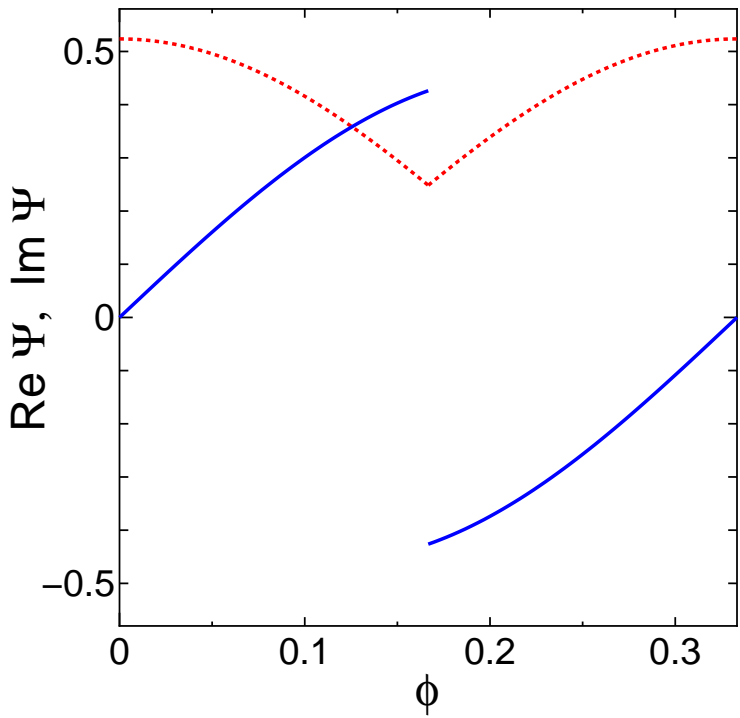}
\end{center}
\caption{Expectation value of the real (dotted line) and imaginary
(solid line) parts of the Polyakov loop at the Roberge-Weiss
transition, $\phi_{RW} = 1/6$.  The values are for $m = m_{dce}$ and $T = T_d$,
but the behavior is qualitatively similar for any Roberge-Weiss transition.}
\label{Fig:Q_re_im} 
\end{figure}

The Roberge-Weiss transition occurs when $\phi_{RW} = 1/6$ 
at sufficiently high temperature for any value of the quark mass.
Then the expectation value of the
imaginary part of the loop is positive to the left of the Roberge-Weiss
transition, and negative to the right.
We illustrate the behavior of the Polyakov loops in Fig. (\ref{Fig:Q_re_im}).
In this figure, $m = m_{dce}$, and we choose a temperature $T_d$.
As expected, the real part of the loop is always nonzero, decreasing
to a minimum at the Roberge-Weiss transition, $\phi_{RW} = 1/6$.  

\begin{figure}[htbp]
\begin{center}
\includegraphics[width=0.6\textwidth]{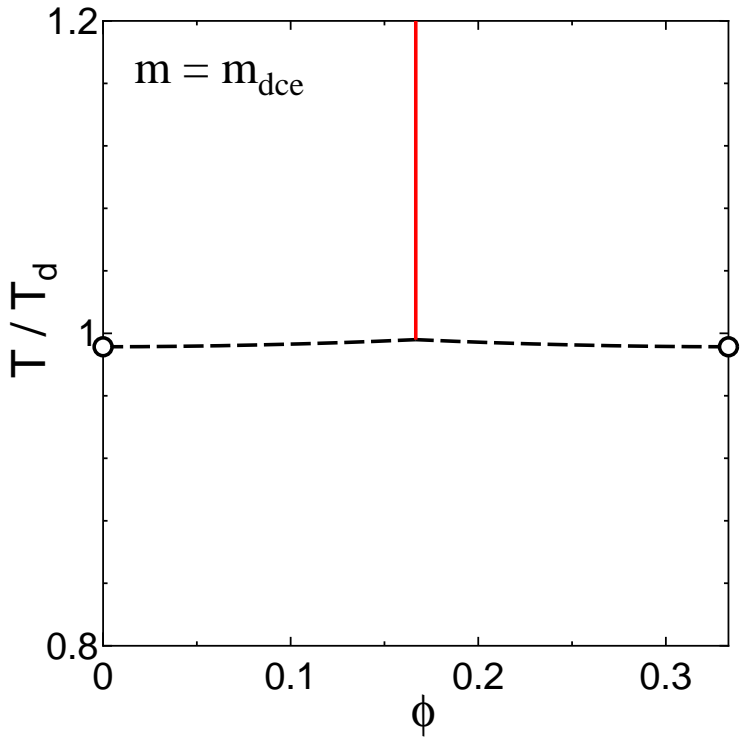}
\end{center}
\caption{Phase diagram at the deconfining critical endpoint, $m = m_{dce}$.
The solid line denotes a Roberge-Weiss transition, of
first order, at $\phi_{RW} = 1/6$;
the dotted lines, first order transitions, which mix deconfining
and Roberge-Weiss transitions.  The points at
$\phi=0$ and $1/3$ are deconfining critical endpoints, and so
of second order.}
\label{Fig:Tdce} 
\end{figure}

The phase diagram 
in the plane of temperature and imaginary chemical potential
is illustrated in Fig. (\ref{Fig:Tdce}).
Here we choose $m = m_{dce}$, so when
$\phi = 0$ (or $1/3$), so there is a second order transition as $T$ 
is varied.  The phase diagram for $m > m_{dce}$ is very similar,
the only difference being that the transition at $\phi = 0$ (or $1/3$)
is of first order instead of second.  

The solid line indicates
the Roberge-Weiss transition at $\phi_{RW} = 1/6$.  
Across this line of first order transitions, by the arguments
in the previous section
the interface tension corresponds directly to the
't Hooft loop.  At a temperature $\approx (1.00 \pm .01) T_d$, 
this line of first order transitions splits into
two lines of first order transitions, drawn as dotted lines.  

The first order transitions for $0< \phi < 1/6$,
and $1/6< \phi < 1/3$, are manifestly those where
the Roberge-Weiss and deconfining transitions mix.  This is clear
above the deconfining critical endpoint,
for $m \geq m_{dce}$, since then the line of first order transitions for
the Roberge-Weiss transition is directly connected by 
lines of first order transitions to the deconfining transition at
$\phi = 0$, where the imaginary part of the Polyakov loop vanishes, and
the transition is entirely one of deconfinement.

As $\phi$ increases from $0$ to
$1/6$, the jump in the imaginary part of the Polyakov loop increases.
In this region, transition is one where the deconfining and Roberge-Weiss
transitions mix.  We have checked that for $\phi \neq 1/6$, that the
jump in the $Z(3)$ phase does not correspond to a $Z(3)$ transformation.
Thus as argued in the previous section, the interface tension across such
first order transitions is not related to a 't Hooft loop.

\begin{figure}[htbp]
\begin{center}
\includegraphics[width=0.6\textwidth]{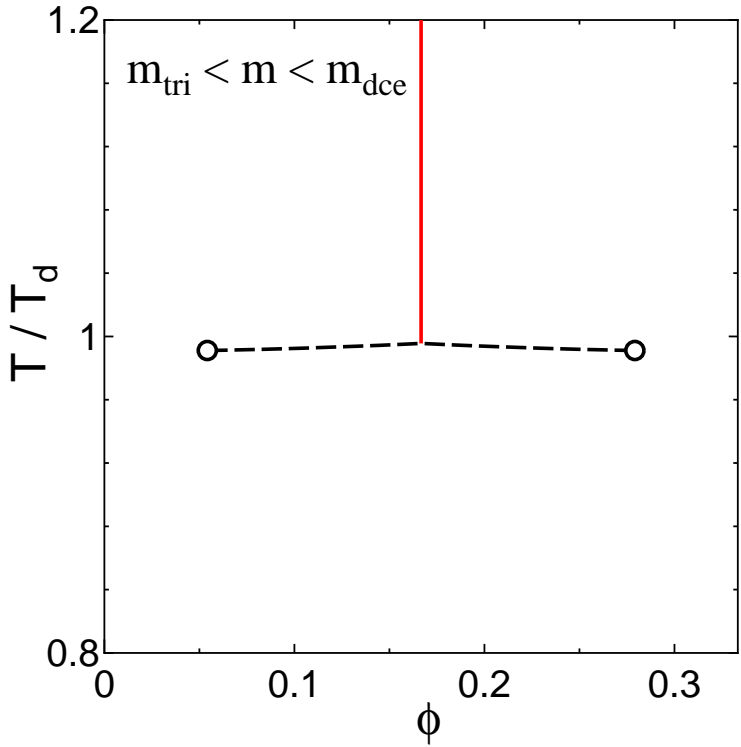}
\end{center}
\caption{Phase diagram for masses below the deconfining critical
endpoint, but above the tri-critical point, $m_{dce} > m > m_{tri}$.
Again, the solid line denotes a Roberge-Weiss transition,
of first order, at $\phi_{RW} = 1/6$;
the dotted lines, first order transitions, which mix deconfining
and Roberge-Weiss transitions.}
\label{Fig:belowTdce} 
\end{figure}

As the quark mass decreases below that for the deconfining critical
endpoint, the two lines of first order transitions for $\phi \neq 1/3$
move closer to $1/6$, see Fig. (\ref{Fig:belowTdce}).  There are two
critical endpoints, at $\phi_c$ and $1/3 - \phi_c$.  At the critical
endpoints, both the real and the imaginary parts of the
Polyakov loop of Eq. (\ref{modified_Polyakov_loop})
are nonzero.  The universality class is that
of $Z(2)$, with the critical field
some combination of real and imaginary parts of the Polyakov loop.

\begin{figure}[htbp]
\begin{center}
\includegraphics[width=0.6\textwidth]{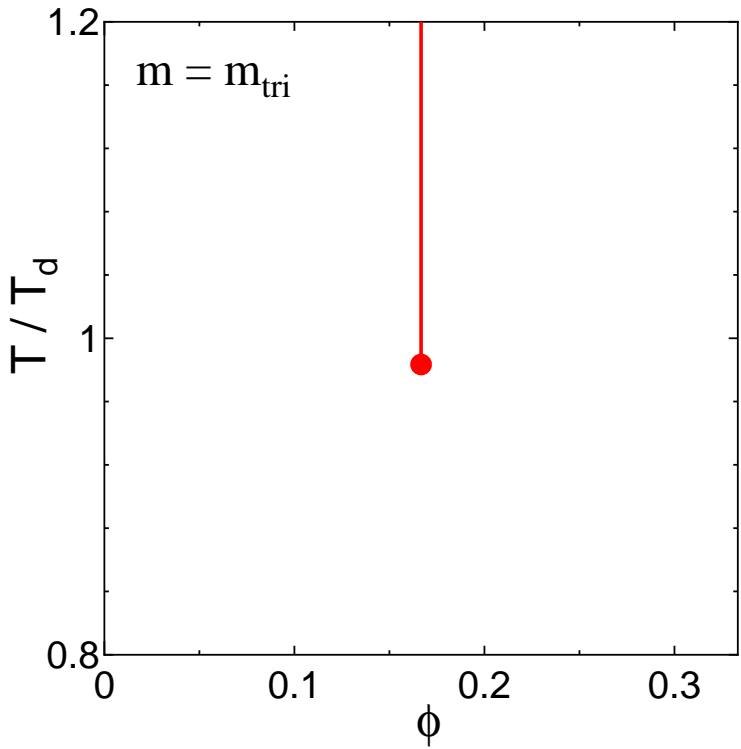}
\end{center}
\caption{Phase diagram for a quark mass at the tri-critical point,
$m = m_{tri}$.  In this region, there is only a Roberge-Weiss transition
at $\phi_{RW} = 1/6$, which ends in a tri-critical point, denoted
by a solid point.  For
$m< m_{tri}$, the line of Roberge-Weiss transitions ends in an
ordinary critical endpoint.}
\label{Fig:Ttri} 
\end{figure}

As $\phi_c \rightarrow 1/6$, the two critical endpoints merge,
and the line of Roberge-Weiss transitions
ends at tri-critical point, as illustrated in Fig. (\ref{Fig:Ttri}).  
In the matrix model, this occurs for a quark mass 
$m_{tri}/T_d \sim 6.3 \pm .1$, at a temperature 
$T_{tri}/T_d \sim 0.983\pm .001$.
At this tri-critical point
the effective theory is that of a single scalar field
in three dimensions, where the mass squared and quartic coupling vanish
at the tri-critical point.   Thus the universality class is mean field,
up to calculable logarithmic corrections.

For masses below $m_{tri}$, there is only a line of Roberge-Weiss
transitions, which end in an ordinary critical endpoint, in the universality
class of the Ising model in three dimensions.  

We note that in general,
the imaginary part of the quark number density is positive for
$0 < \phi < 1/6$, and negative for $1/6 < \phi < 1/3$.  For temperatures
sufficiently high that there is a Roberge-Weiss transition, the sign
of the imaginary part of the quark number density flips at $\phi_{RW} = 1/6$,
and there is a first order transition.  For temperatures below that
for the Roberge-Weiss transition, the imaginary part of the quark number
density is still positive for 
$0 < \phi < 1/6$, and negative for $1/6 < \phi < 1/3$, but because
there is no phase transition, it vanishes for $\phi_{RW} = 1/6$.

\subsubsection{Thermodynamics of Roberge-Weiss transitions}

We now turn to the thermodynamics of Roberge-Weiss transitions.
We concentrate on the transitions for $\phi_{RW} = 1/6$, as those
for $\phi \neq 1/6$ are qualitatively similar.

\begin{figure}[htbp]
\begin{center}
\includegraphics[width=0.45\textwidth]{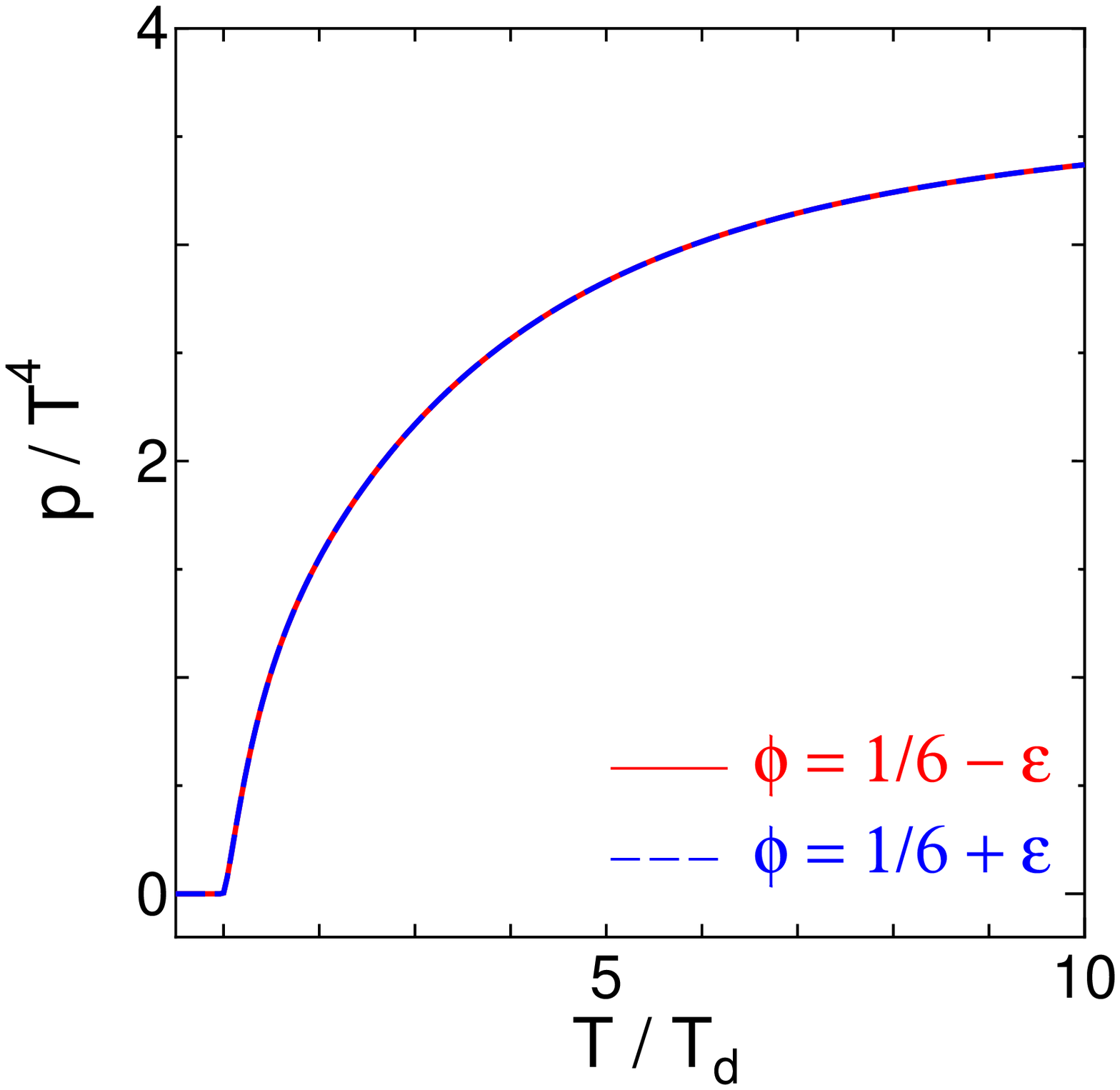}
\includegraphics[width=0.45\textwidth]{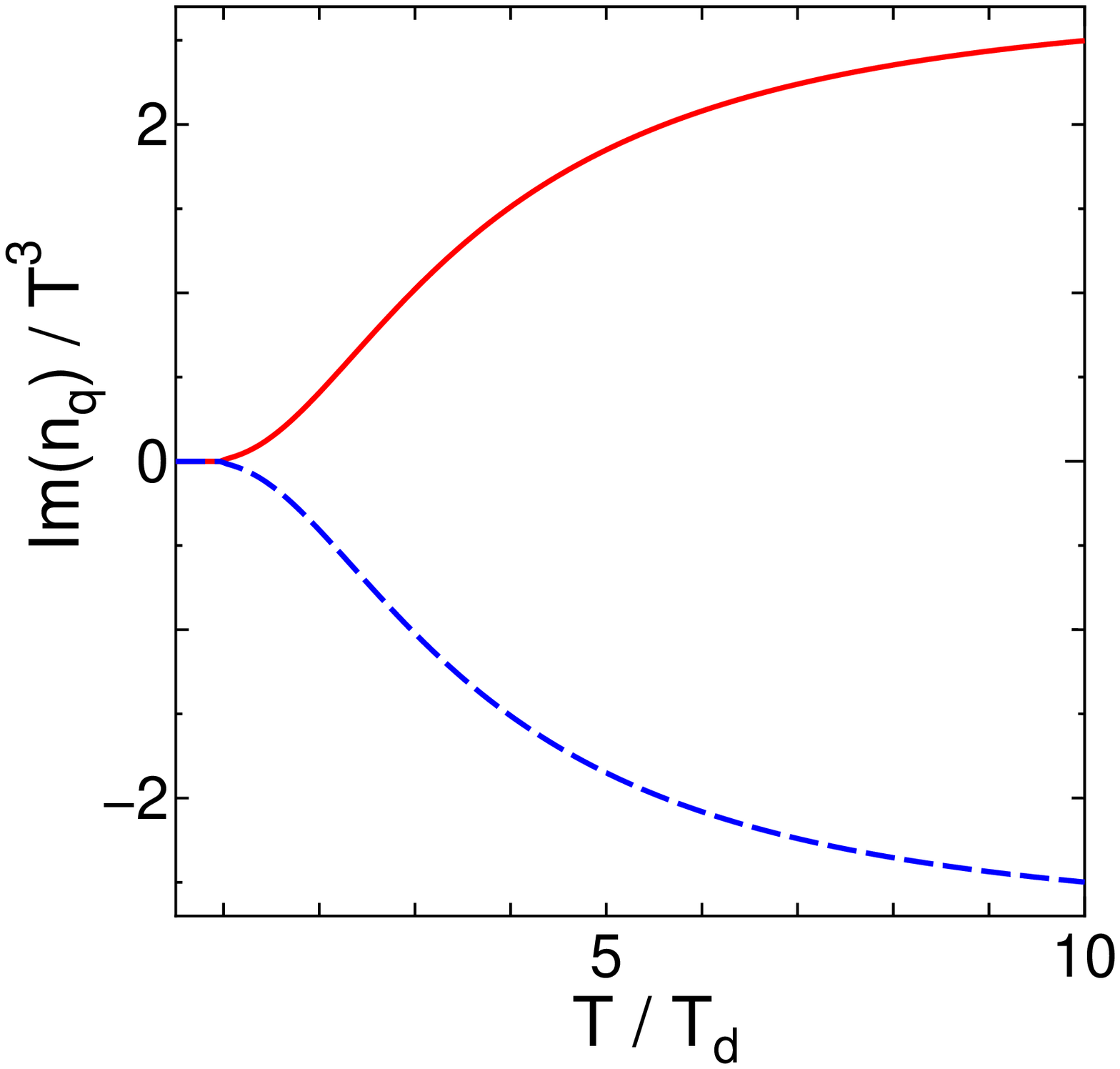} 
\end{center}
\caption{The temperature dependence of the pressure and 
the imaginary part of the quark number density
across the Roberge-Weiss transition at $\phi_{RW} = 1/6$
with $m = m_{dce}$.
All quantities are scaled by appropriate powers of the temperature 
to be dimensionless, and shown as functions of $T/T_d$.
The solid and dashed lines represents $\phi=1/6 - \epsilon$ and
$\phi=1/6+\epsilon$, respectively.}
\label{Fig:TQa} 
\end{figure}

\begin{figure}[htbp]
\begin{center}
\includegraphics[width=0.45\textwidth]{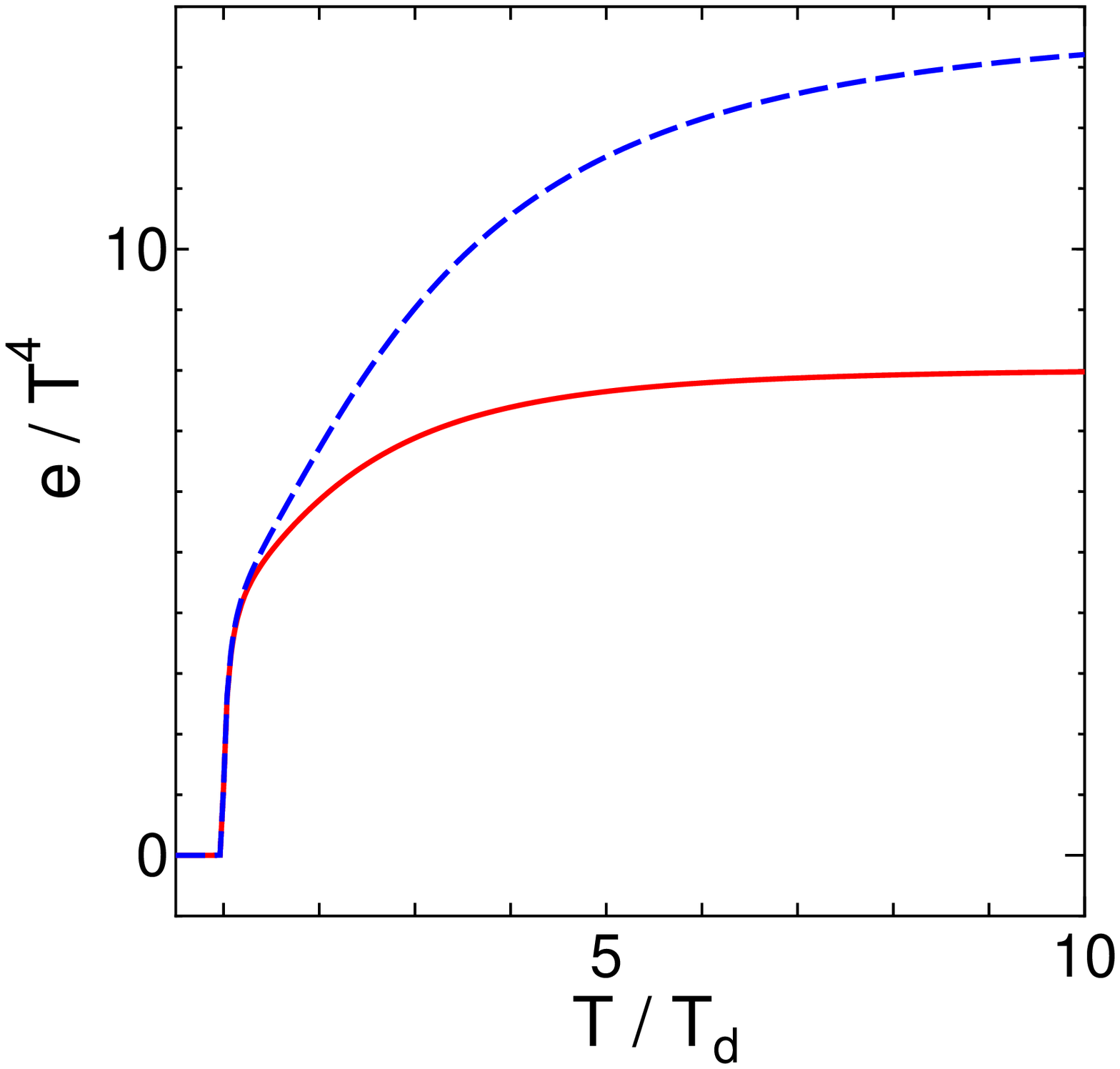}
\includegraphics[width=0.45\textwidth]{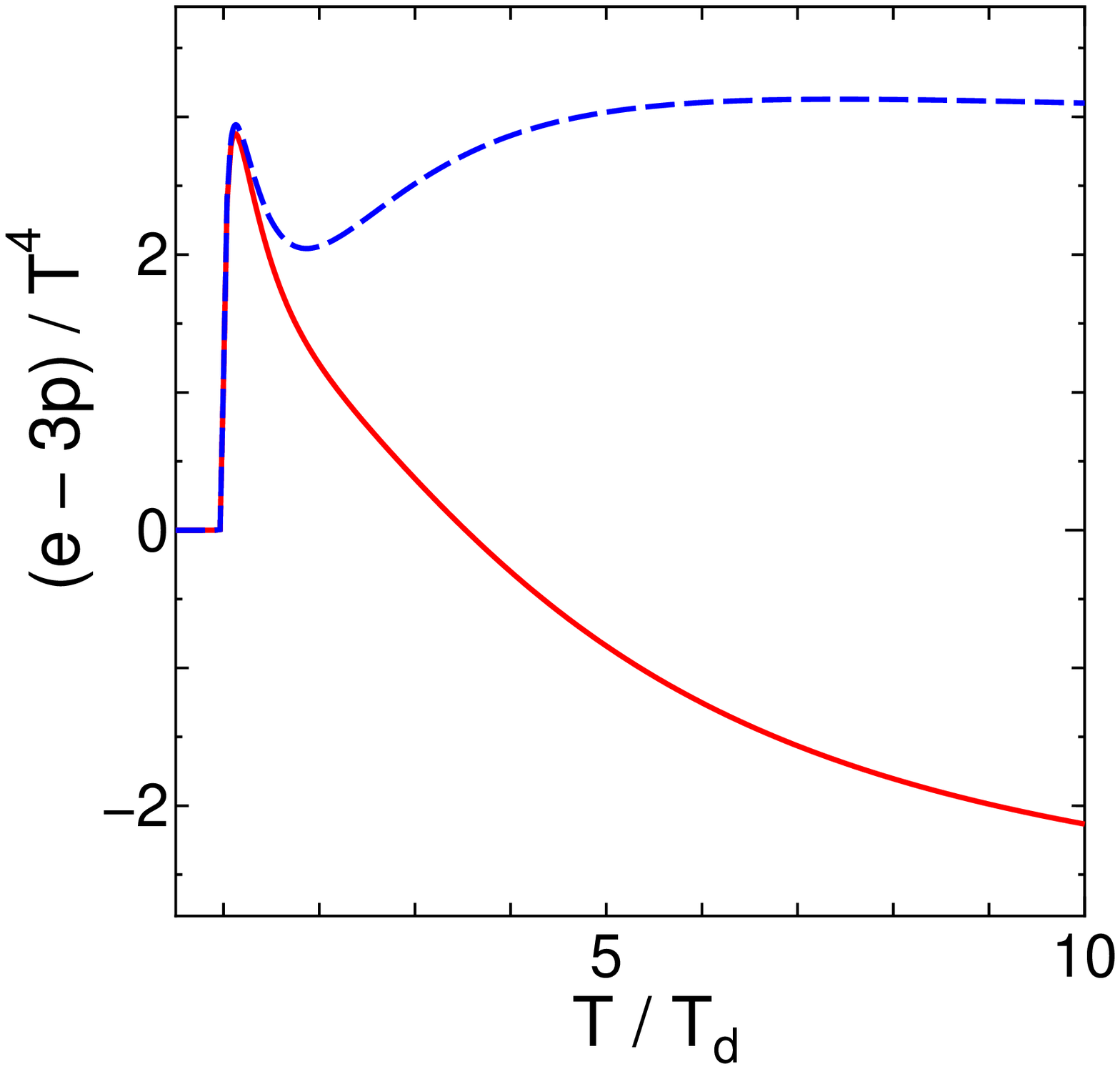} 
\end{center}
\caption{The temperature dependence of the energy density,
$e(T)/T^4$ and 
the interaction measure, $(e-3p)/T^4$,
across the Roberge-Weiss transition at $\phi_{RW} = 1/6$,
with $m = m_{dce}$.
The solid and dashed lines represents $\phi=1/6 - \epsilon$ and
$\phi=1/6+\epsilon$, respectively.}
\label{Fig:TQb} 
\end{figure}

We begin with a quark mass at the deconfining critical endpoint 
(when $\phi = 0$).
In Fig. (\ref{Fig:TQa}) we show the pressure, $p(T)$, the
imaginary part of the quark number density, $n_q(T)$.
Both quantities are rescaled by powers of the temperature to
be dimensionless, so we show $p(T)/T^4$ and ${\rm Im}~n_q(T)/T^3$.
Similarly, in Fig. (\ref{Fig:TQb}) we show
the rescaled energy density $e(T)/T^4$, 
Eq. (\ref{energy_density}), and the interaction measure,
$(e - 3p)/T^4$.

Fig. (\ref{Fig:TQa}) shows that the pressure is like the real part of 
the Polyakov loop, and is continuous across the Roberge-Weiss transition.
In contrast, the imaginary part of the quark number density is like
the imaginary part of the Polyakov loop, and flips sign across this
transition.

Because the pressure is continuous, the change in the
imaginary part of the quark number density implies that the
energy density is discontinuous across the transition.  Thus
this change in the imaginary part of the quark number density is
the only reason the transition is of first order, Fig. (\ref{Fig:TQb}).
On the right hand side of the transition, for $\phi_{RW} = 1/6^+$,
the smaller value of the energy density implies that the interaction
measure eventually becomes negative, Fig. (\ref{Fig:TQb}),
for temperatures above $\sim 3 \, T_d$.  

To the left of the Roberge-Weiss transition, the interaction measure
displays a characteristic two peak structure \cite{Kashiwa:2012va}. 
The peak near $T_d$ is due to the gluons, that at several times
$T_d$ is due to the quarks, because they are so heavy.  Because
the energy density to the right of the Roberge-Weiss transition is
negative, the peak in the interaction measure persists, but that
due to the quarks is completely washed out by the negative energy
density for $\phi_{RW} = 1/6^+$.
At high temperatures, the interaction measure has the same value for
$\phi_{RW} = 1/6^\pm$, but with opposite sign, see Sec. \ref{sec:high}.

\begin{figure}[htbp]
\begin{center}
\includegraphics[width=0.45\textwidth]{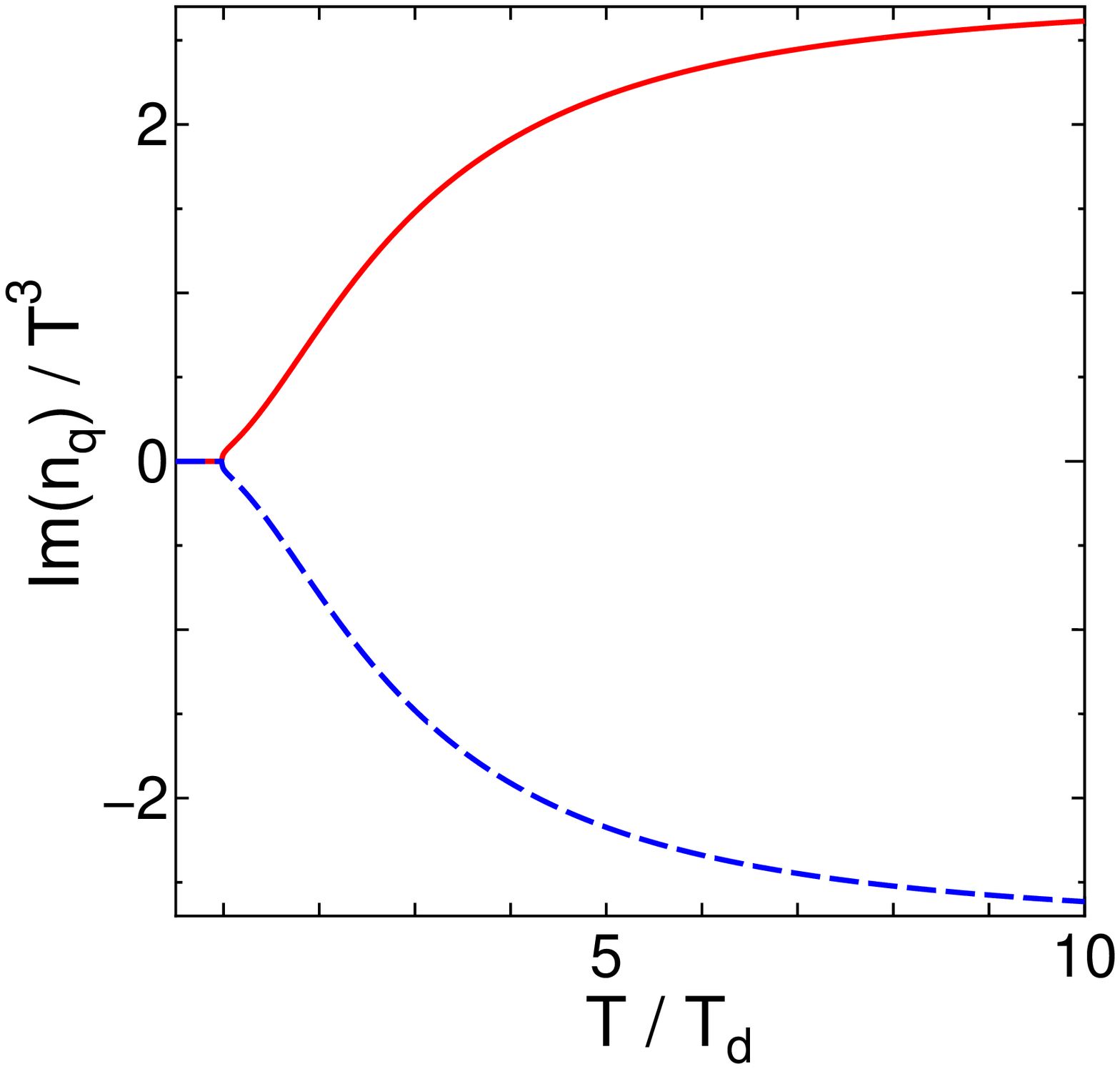}
\includegraphics[width=0.45\textwidth]{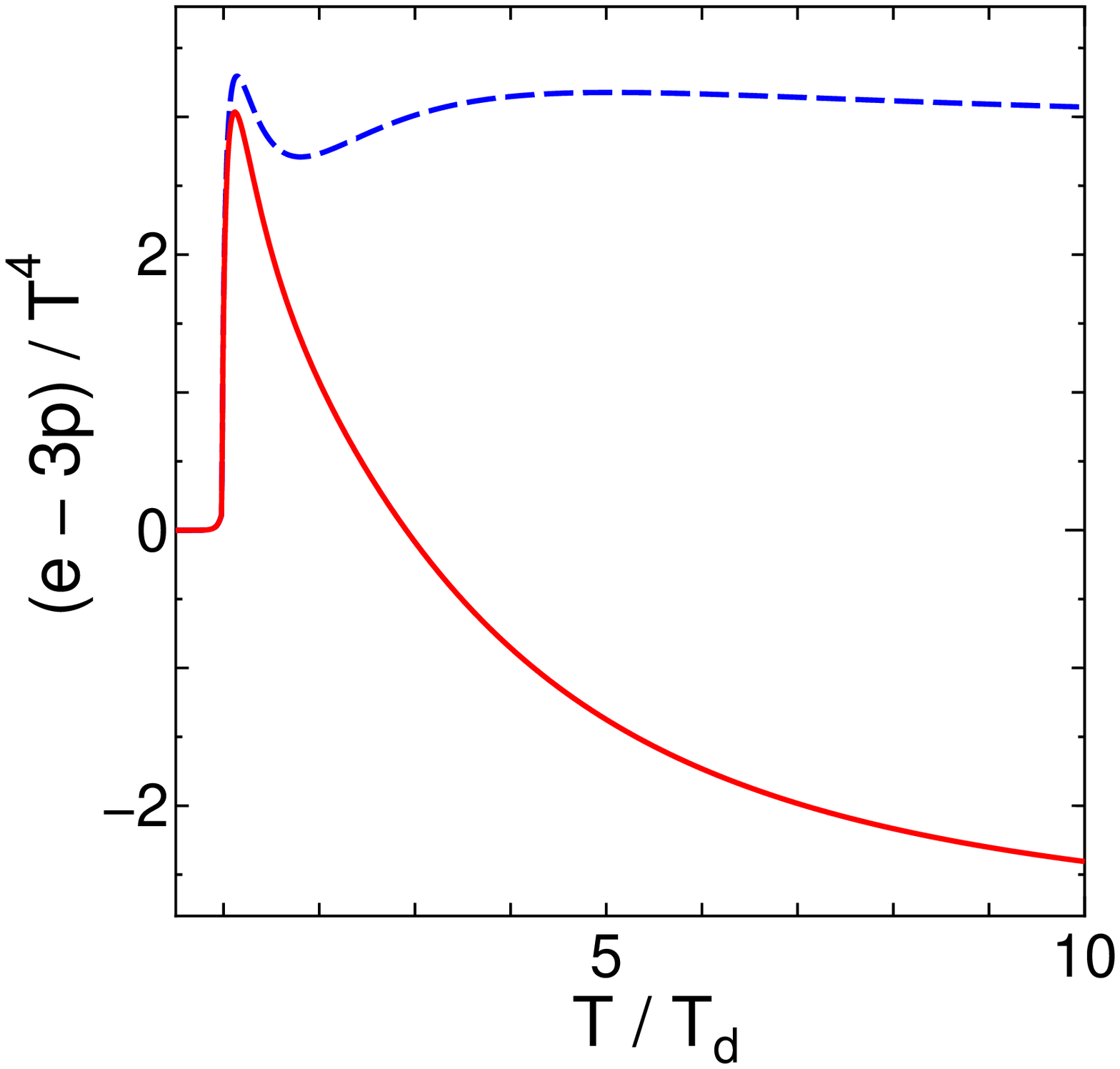} 
\end{center}
\caption{The temperature dependence of the imaginary part of the
quark number density, ${\rm Im}(n_q)/T^3$, and 
the interaction measure, $(e-3p)/T^4$,
across the Roberge-Weiss transition at $\phi_{RW} = 1/6$,
with $m = m_{tri}$.
The solid and dashed lines represents $\phi=1/6 - \epsilon$ and
$\phi=1/6+\epsilon$, respectively.}
\label{Fig:tri} 
\end{figure}

The behavior of the thermodynamic quantities is similar across
the Roberge-Weiss transition for other quark masses.  
In Fig. (\ref{Fig:tri}) we show the behavior of the imaginary
part of the quark number density, and the interaction measure,
for $m=m_{tri}$.
This is similar to the
behavior at the deconfining critical endpoint, Figs.
(\ref{Fig:TQa}) and (\ref{Fig:TQb}), and so we do not bother
to show the corresponding results for the pressure and the
energy density.

\begin{figure}[htbp]
\begin{center}
\includegraphics[width=0.6\textwidth]{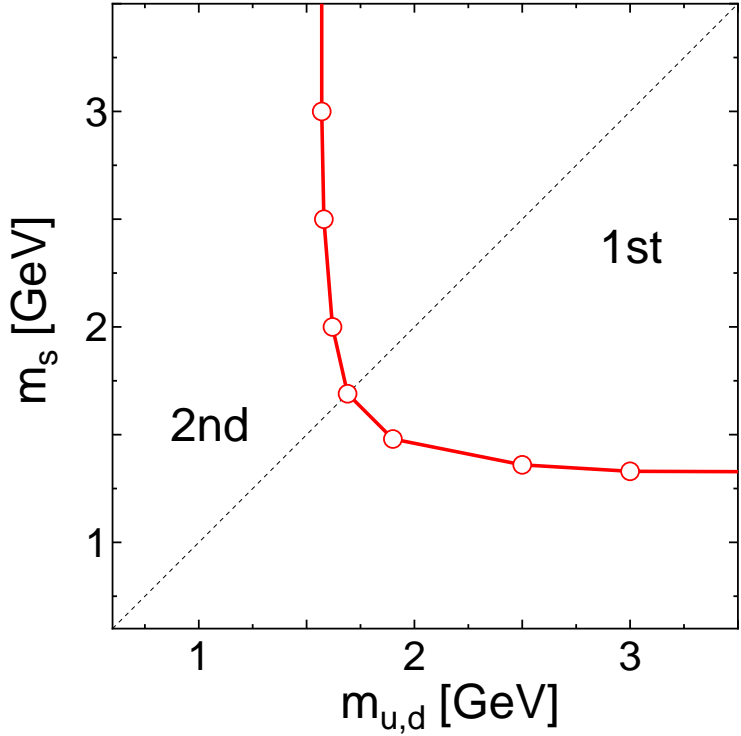} 
\end{center}
\caption{The dependence of the tri-critical endpoint for $2+1$ flavors.}
\label{Fig:CP}
\end{figure}

One can compute the position of the tri-critical point versus the
number of flavors, and is illustrated in Fig. (\ref{Fig:CP}). 
Recent lattice QCD data~\cite{Bonati:2011a, *Bonati:2011b,
*Wu:2013bfa} suggest that there is the tri-critical line 
in the heavy quark mass region at the RW endpoint.
The matrix model reproduces the result and 
values of $m_{tri}/T_{tri}$ are
summarized in Table. \ref{Table}.
\begin{table}[h]
\begin{center}
\begin{tabular}{cccc} \\ 
\hline \hline
&~~$m_l = \infty$~~  &  ~~$m_l = m_s$~~ & ~~$m_s = \infty$~~ \\ 
\hline
~~Ref.~\cite{Fromm:2012yg} ~~ &  5.56(3) &  6.66(3)
&   6.25(3)      
\\
\hline
~~This model~~ &  5.0  &  6.4 &  5.9
\\ 
\hline \hline
\end{tabular} \caption{
Summary for $m_{tri} / T_{tri}$ in the recent LQCD simulation with the strong
coupling and hopping parameter expansions~\cite{Fromm:2012yg} and the 
matrix model for deconfinement. 
We show our result down to first decimal place.}
\label{Table} \end{center} \end{table}
These values well reproduce the recent LQCD simulation with the strong coupling
and hopping parameter expansions~\cite{Fromm:2012yg}.

\subsection{Summary}

In this paper we showed that
't Hooft loops of arbitrary $Z(N)$ charge are well defined even
with dynamical quarks at a Roberge-Weiss transition, for 
$\phi_{RW} = k/(2N)$.
To leading order in weak coupling, 
the 't Hooft loop satisfies Casimir scaling
in the pure glue theory, but not with quarks.

For three colors
we computed thermodynamic behavior at large quark mass using
an effective matrix model for deconfinement, and computed the
form of the Columbia plot.  We computed the interaction measure 
about the Roberge-Weiss transition, $\phi_{RW} = 1/6 \pm \epsilon$, 
and find an enhancement on one side of the transition, but not the other.

Considering an imaginary 
chemical potential is clearly useful to discriminate between various effective
models of deconfinement.  The relationship to the 't Hooft loop suggests
that it probes more fundamental aspects of the dynamics in unforeseen
ways.  

\noindent \begin{acknowledgments}
K.K. is supported by the RIKEN
Special Postdoctoral Researchers Program. 
The research of R.D.P. is supported by the U.S. Department of Energy under
contract \#DE-AC02-98CH10886.  We also thank Chris
Korthals-Altes and Philippe de Forcrand for 
numerous helpful discussions.
\end{acknowledgments}

\bibliography{RW}

\begin{thebibliography}{63}%
\makeatletter
\providecommand \@ifxundefined [1]{%
 \@ifx{#1\undefined}
}%
\providecommand \@ifnum [1]{%
 \ifnum #1\expandafter \@firstoftwo
 \else \expandafter \@secondoftwo
 \fi
}%
\providecommand \@ifx [1]{%
 \ifx #1\expandafter \@firstoftwo
 \else \expandafter \@secondoftwo
 \fi
}%
\providecommand \natexlab [1]{#1}%
\providecommand \enquote  [1]{``#1''}%
\providecommand \bibnamefont  [1]{#1}%
\providecommand \bibfnamefont [1]{#1}%
\providecommand \citenamefont [1]{#1}%
\providecommand \href@noop [0]{\@secondoftwo}%
\providecommand \href [0]{\begingroup \@sanitize@url \@href}%
\providecommand \@href[1]{\@@startlink{#1}\@@href}%
\providecommand \@@href[1]{\endgroup#1\@@endlink}%
\providecommand \@sanitize@url [0]{\catcode `\\12\catcode `\$12\catcode
  `\&12\catcode `\#12\catcode `\^12\catcode `\_12\catcode `\%12\relax}%
\providecommand \@@startlink[1]{}%
\providecommand \@@endlink[0]{}%
\providecommand \url  [0]{\begingroup\@sanitize@url \@url }%
\providecommand \@url [1]{\endgroup\@href {#1}{\urlprefix }}%
\providecommand \urlprefix  [0]{URL }%
\providecommand \Eprint [0]{\href }%
\providecommand \doibase [0]{http://dx.doi.org/}%
\providecommand \selectlanguage [0]{\@gobble}%
\providecommand \bibinfo  [0]{\@secondoftwo}%
\providecommand \bibfield  [0]{\@secondoftwo}%
\providecommand \translation [1]{[#1]}%
\providecommand \BibitemOpen [0]{}%
\providecommand \bibitemStop [0]{}%
\providecommand \bibitemNoStop [0]{.\EOS\space}%
\providecommand \EOS [0]{\spacefactor3000\relax}%
\providecommand \BibitemShut  [1]{\csname bibitem#1\endcsname}%
\let\auto@bib@innerbib\@empty
\bibitem [{\citenamefont {de~Forcrand}(2009)}]{Forcrand:2009}%
  \BibitemOpen
  \bibfield  {author} {\bibinfo {author} {\bibfnamefont {P.}~\bibnamefont
  {de~Forcrand}},\ }\href@noop {} {\bibfield  {journal} {\bibinfo  {journal}
  {PoS (LAT2009)}\ ,\ \bibinfo {pages} {010}} (\bibinfo {year} {2009})},\
  \Eprint {http://arxiv.org/abs/1005.0539} {arXiv:1005.0539 [hep-lat]}
  \BibitemShut {NoStop}%
\bibitem [{\citenamefont {Roberge}\ and\ \citenamefont
  {Weiss}(1986)}]{Roberge:1986}%
  \BibitemOpen
  \bibfield  {author} {\bibinfo {author} {\bibfnamefont {A.}~\bibnamefont
  {Roberge}}\ and\ \bibinfo {author} {\bibfnamefont {N.}~\bibnamefont
  {Weiss}},\ }\href {\doibase 10.1016/0550-3213(86)90582-1} {\bibfield
  {journal} {\bibinfo  {journal} {Nucl. Phys.}\ }\textbf {\bibinfo {volume}
  {B275}},\ \bibinfo {pages} {734} (\bibinfo {year} {1986})}\BibitemShut
  {NoStop}%
\bibitem [{\citenamefont {Alford}\ \emph {et~al.}(1999)\citenamefont {Alford},
  \citenamefont {Kapustin},\ and\ \citenamefont {Wilczek}}]{Alford:1998sd}%
  \BibitemOpen
  \bibfield  {author} {\bibinfo {author} {\bibfnamefont {M.~G.}\ \bibnamefont
  {Alford}}, \bibinfo {author} {\bibfnamefont {A.}~\bibnamefont {Kapustin}}, \
  and\ \bibinfo {author} {\bibfnamefont {F.}~\bibnamefont {Wilczek}},\ }\href
  {\doibase 10.1103/PhysRevD.59.054502} {\bibfield  {journal} {\bibinfo
  {journal} {Phys.Rev.}\ }\textbf {\bibinfo {volume} {D59}},\ \bibinfo {pages}
  {054502} (\bibinfo {year} {1999})},\ \Eprint
  {http://arxiv.org/abs/hep-lat/9807039} {arXiv:hep-lat/9807039 [hep-lat]}
  \BibitemShut {NoStop}%
\bibitem [{\citenamefont {Hart}\ \emph {et~al.}(2001)\citenamefont {Hart},
  \citenamefont {Laine},\ and\ \citenamefont {Philipsen}}]{Hart:2000ef}%
  \BibitemOpen
  \bibfield  {author} {\bibinfo {author} {\bibfnamefont {A.}~\bibnamefont
  {Hart}}, \bibinfo {author} {\bibfnamefont {M.}~\bibnamefont {Laine}}, \ and\
  \bibinfo {author} {\bibfnamefont {O.}~\bibnamefont {Philipsen}},\ }\href
  {\doibase 10.1016/S0370-2693(01)00355-0} {\bibfield  {journal} {\bibinfo
  {journal} {Phys.Lett.}\ }\textbf {\bibinfo {volume} {B505}},\ \bibinfo
  {pages} {141} (\bibinfo {year} {2001})},\ \Eprint
  {http://arxiv.org/abs/hep-lat/0010008} {arXiv:hep-lat/0010008 [hep-lat]}
  \BibitemShut {NoStop}%
\bibitem [{\citenamefont {de~Forcrand}\ and\ \citenamefont
  {Philipsen}(2002)}]{deForcrand:2002ci}%
  \BibitemOpen
  \bibfield  {author} {\bibinfo {author} {\bibfnamefont {P.}~\bibnamefont
  {de~Forcrand}}\ and\ \bibinfo {author} {\bibfnamefont {O.}~\bibnamefont
  {Philipsen}},\ }\href {\doibase 10.1016/S0550-3213(02)00626-0} {\bibfield
  {journal} {\bibinfo  {journal} {Nucl.Phys.}\ }\textbf {\bibinfo {volume}
  {B642}},\ \bibinfo {pages} {290} (\bibinfo {year} {2002})},\ \Eprint
  {http://arxiv.org/abs/hep-lat/0205016} {arXiv:hep-lat/0205016 [hep-lat]}
  \BibitemShut {NoStop}%
\bibitem [{\citenamefont {D'Elia}\ and\ \citenamefont
  {Lombardo}(2003)}]{D'Elia:2002gd}%
  \BibitemOpen
  \bibfield  {author} {\bibinfo {author} {\bibfnamefont {M.}~\bibnamefont
  {D'Elia}}\ and\ \bibinfo {author} {\bibfnamefont {M.-P.}\ \bibnamefont
  {Lombardo}},\ }\href {\doibase 10.1103/PhysRevD.67.014505} {\bibfield
  {journal} {\bibinfo  {journal} {Phys.Rev.}\ }\textbf {\bibinfo {volume}
  {D67}},\ \bibinfo {pages} {014505} (\bibinfo {year} {2003})},\ \Eprint
  {http://arxiv.org/abs/hep-lat/0209146} {arXiv:hep-lat/0209146 [hep-lat]}
  \BibitemShut {NoStop}%
\bibitem [{\citenamefont {D'Elia}\ and\ \citenamefont
  {Lombardo}(2004)}]{D'Elia:2004at}%
  \BibitemOpen
  \bibfield  {author} {\bibinfo {author} {\bibfnamefont {M.}~\bibnamefont
  {D'Elia}}\ and\ \bibinfo {author} {\bibfnamefont {M.~P.}\ \bibnamefont
  {Lombardo}},\ }\href {\doibase 10.1103/PhysRevD.70.074509} {\bibfield
  {journal} {\bibinfo  {journal} {Phys.Rev.}\ }\textbf {\bibinfo {volume}
  {D70}},\ \bibinfo {pages} {074509} (\bibinfo {year} {2004})},\ \Eprint
  {http://arxiv.org/abs/hep-lat/0406012} {arXiv:hep-lat/0406012 [hep-lat]}
  \BibitemShut {NoStop}%
\bibitem [{\citenamefont {Chen}\ and\ \citenamefont {Luo}(2005)}]{Chen:2004tb}%
  \BibitemOpen
  \bibfield  {author} {\bibinfo {author} {\bibfnamefont {H.-S.}\ \bibnamefont
  {Chen}}\ and\ \bibinfo {author} {\bibfnamefont {X.-Q.}\ \bibnamefont {Luo}},\
  }\href {\doibase 10.1103/PhysRevD.72.034504} {\bibfield  {journal} {\bibinfo
  {journal} {Phys.Rev.}\ }\textbf {\bibinfo {volume} {D72}},\ \bibinfo {pages}
  {034504} (\bibinfo {year} {2005})},\ \Eprint
  {http://arxiv.org/abs/hep-lat/0411023} {arXiv:hep-lat/0411023 [hep-lat]}
  \BibitemShut {NoStop}%
\bibitem [{\citenamefont {de~Forcrand}\ and\ \citenamefont
  {Philipsen}(2007)}]{deForcrand:2006pv}%
  \BibitemOpen
  \bibfield  {author} {\bibinfo {author} {\bibfnamefont {P.}~\bibnamefont
  {de~Forcrand}}\ and\ \bibinfo {author} {\bibfnamefont {O.}~\bibnamefont
  {Philipsen}},\ }\href {\doibase 10.1088/1126-6708/2007/01/077} {\bibfield
  {journal} {\bibinfo  {journal} {JHEP}\ }\textbf {\bibinfo {volume} {0701}},\
  \bibinfo {pages} {077} (\bibinfo {year} {2007})},\ \Eprint
  {http://arxiv.org/abs/hep-lat/0607017} {arXiv:hep-lat/0607017 [hep-lat]}
  \BibitemShut {NoStop}%
\bibitem [{\citenamefont {Wu}\ \emph {et~al.}(2007)\citenamefont {Wu},
  \citenamefont {Luo},\ and\ \citenamefont {Chen}}]{Wu:2006su}%
  \BibitemOpen
  \bibfield  {author} {\bibinfo {author} {\bibfnamefont {L.-K.}\ \bibnamefont
  {Wu}}, \bibinfo {author} {\bibfnamefont {X.-Q.}\ \bibnamefont {Luo}}, \ and\
  \bibinfo {author} {\bibfnamefont {H.-S.}\ \bibnamefont {Chen}},\ }\href
  {\doibase 10.1103/PhysRevD.76.034505} {\bibfield  {journal} {\bibinfo
  {journal} {Phys.Rev.}\ }\textbf {\bibinfo {volume} {D76}},\ \bibinfo {pages}
  {034505} (\bibinfo {year} {2007})},\ \Eprint
  {http://arxiv.org/abs/hep-lat/0611035} {arXiv:hep-lat/0611035 [hep-lat]}
  \BibitemShut {NoStop}%
\bibitem [{\citenamefont {D'Elia}\ \emph {et~al.}(2007)\citenamefont {D'Elia},
  \citenamefont {Di~Renzo},\ and\ \citenamefont {Lombardo}}]{D'Elia:2007ke}%
  \BibitemOpen
  \bibfield  {author} {\bibinfo {author} {\bibfnamefont {M.}~\bibnamefont
  {D'Elia}}, \bibinfo {author} {\bibfnamefont {F.}~\bibnamefont {Di~Renzo}}, \
  and\ \bibinfo {author} {\bibfnamefont {M.~P.}\ \bibnamefont {Lombardo}},\
  }\href {\doibase 10.1103/PhysRevD.76.114509} {\bibfield  {journal} {\bibinfo
  {journal} {Phys.Rev.}\ }\textbf {\bibinfo {volume} {D76}},\ \bibinfo {pages}
  {114509} (\bibinfo {year} {2007})},\ \Eprint {http://arxiv.org/abs/0705.3814}
  {arXiv:0705.3814 [hep-lat]} \BibitemShut {NoStop}%
\bibitem [{\citenamefont {de~Forcrand}\ and\ \citenamefont
  {Philipsen}(2008)}]{deForcrand:2008vr}%
  \BibitemOpen
  \bibfield  {author} {\bibinfo {author} {\bibfnamefont {P.}~\bibnamefont
  {de~Forcrand}}\ and\ \bibinfo {author} {\bibfnamefont {O.}~\bibnamefont
  {Philipsen}},\ }\href {\doibase 10.1088/1126-6708/2008/11/012} {\bibfield
  {journal} {\bibinfo  {journal} {JHEP}\ }\textbf {\bibinfo {volume} {0811}},\
  \bibinfo {pages} {012} (\bibinfo {year} {2008})},\ \Eprint
  {http://arxiv.org/abs/0808.1096} {arXiv:0808.1096 [hep-lat]} \BibitemShut
  {NoStop}%
\bibitem [{\citenamefont {D'Elia}\ and\ \citenamefont
  {Sanfilippo}(2009{\natexlab{a}})}]{D'Elia:2009tm}%
  \BibitemOpen
  \bibfield  {author} {\bibinfo {author} {\bibfnamefont {M.}~\bibnamefont
  {D'Elia}}\ and\ \bibinfo {author} {\bibfnamefont {F.}~\bibnamefont
  {Sanfilippo}},\ }\href {\doibase 10.1103/PhysRevD.80.014502} {\bibfield
  {journal} {\bibinfo  {journal} {Phys.Rev.}\ }\textbf {\bibinfo {volume}
  {D80}},\ \bibinfo {pages} {014502} (\bibinfo {year} {2009}{\natexlab{a}})},\
  \Eprint {http://arxiv.org/abs/0904.1400} {arXiv:0904.1400 [hep-lat]}
  \BibitemShut {NoStop}%
\bibitem [{\citenamefont {D'Elia}\ and\ \citenamefont
  {Sanfilippo}(2009{\natexlab{b}})}]{D'Elia:2009qz}%
  \BibitemOpen
  \bibfield  {author} {\bibinfo {author} {\bibfnamefont {M.}~\bibnamefont
  {D'Elia}}\ and\ \bibinfo {author} {\bibfnamefont {F.}~\bibnamefont
  {Sanfilippo}},\ }\href {\doibase 10.1103/PhysRevD.80.111501} {\bibfield
  {journal} {\bibinfo  {journal} {Phys.Rev.}\ }\textbf {\bibinfo {volume}
  {D80}},\ \bibinfo {pages} {111501} (\bibinfo {year} {2009}{\natexlab{b}})},\
  \Eprint {http://arxiv.org/abs/0909.0254} {arXiv:0909.0254 [hep-lat]}
  \BibitemShut {NoStop}%
\bibitem [{\citenamefont {de~Forcrand}\ and\ \citenamefont
  {Philipsen}(2010)}]{deForcrand:2010he}%
  \BibitemOpen
  \bibfield  {author} {\bibinfo {author} {\bibfnamefont {P.}~\bibnamefont
  {de~Forcrand}}\ and\ \bibinfo {author} {\bibfnamefont {O.}~\bibnamefont
  {Philipsen}},\ }\href {\doibase 10.1103/PhysRevLett.105.152001} {\bibfield
  {journal} {\bibinfo  {journal} {Phys.Rev.Lett.}\ }\textbf {\bibinfo {volume}
  {105}},\ \bibinfo {pages} {152001} (\bibinfo {year} {2010})},\ \Eprint
  {http://arxiv.org/abs/1004.3144} {arXiv:1004.3144 [hep-lat]} \BibitemShut
  {NoStop}%
\bibitem [{\citenamefont {Cea}\ \emph {et~al.}(2012)\citenamefont {Cea},
  \citenamefont {Cosmai}, \citenamefont {D'Elia}, \citenamefont {Papa},\ and\
  \citenamefont {Sanfilippo}}]{Cea:2012ev}%
  \BibitemOpen
  \bibfield  {author} {\bibinfo {author} {\bibfnamefont {P.}~\bibnamefont
  {Cea}}, \bibinfo {author} {\bibfnamefont {L.}~\bibnamefont {Cosmai}},
  \bibinfo {author} {\bibfnamefont {M.}~\bibnamefont {D'Elia}}, \bibinfo
  {author} {\bibfnamefont {A.}~\bibnamefont {Papa}}, \ and\ \bibinfo {author}
  {\bibfnamefont {F.}~\bibnamefont {Sanfilippo}},\ }\href {\doibase
  10.1103/PhysRevD.85.094512} {\bibfield  {journal} {\bibinfo  {journal}
  {Phys.Rev.}\ }\textbf {\bibinfo {volume} {D85}},\ \bibinfo {pages} {094512}
  (\bibinfo {year} {2012})},\ \Eprint {http://arxiv.org/abs/1202.5700}
  {arXiv:1202.5700 [hep-lat]} \BibitemShut {NoStop}%
\bibitem [{\citenamefont {Bonati}\ \emph {et~al.}(2011)\citenamefont {Bonati},
  \citenamefont {Cossu}, \citenamefont {D'Elia},\ and\ \citenamefont
  {Sanfilippo}}]{Bonati:2011a}%
  \BibitemOpen
  \bibfield  {author} {\bibinfo {author} {\bibfnamefont {C.}~\bibnamefont
  {Bonati}}, \bibinfo {author} {\bibfnamefont {G.}~\bibnamefont {Cossu}},
  \bibinfo {author} {\bibfnamefont {M.}~\bibnamefont {D'Elia}}, \ and\ \bibinfo
  {author} {\bibfnamefont {F.}~\bibnamefont {Sanfilippo}},\ }\href {\doibase
  10.1103/PhysRevD.83.054505} {\bibfield  {journal} {\bibinfo  {journal} {Phys.
  Rev. D}\ }\textbf {\bibinfo {volume} {83}},\ \bibinfo {pages} {054505}
  (\bibinfo {year} {2011})},\ \Eprint {http://arxiv.org/abs/1011.4515}
  {arXiv:1011.4515 [hep-lat]} \BibitemShut {NoStop}%
\bibitem [{\citenamefont {Bonati}\ \emph {et~al.}(2012)\citenamefont {Bonati},
  \citenamefont {de~Forcrand}, \citenamefont {Cossu}, \citenamefont {D'Elia},
  \citenamefont {Philipsen},\ and\ \citenamefont {Sanfilippo}}]{Bonati:2011b}%
  \BibitemOpen
  \bibfield  {author} {\bibinfo {author} {\bibfnamefont {C.}~\bibnamefont
  {Bonati}}, \bibinfo {author} {\bibfnamefont {P.}~\bibnamefont {de~Forcrand}},
  \bibinfo {author} {\bibfnamefont {G.}~\bibnamefont {Cossu}}, \bibinfo
  {author} {\bibfnamefont {M.}~\bibnamefont {D'Elia}}, \bibinfo {author}
  {\bibfnamefont {O.}~\bibnamefont {Philipsen}}, \ and\ \bibinfo {author}
  {\bibfnamefont {F.}~\bibnamefont {Sanfilippo}},\ }\href@noop {} {\  (\bibinfo
  {year} {2012})},\ \Eprint {http://arxiv.org/abs/1201.2769} {arXiv:1201.2769
  [hep-lat]} \BibitemShut {NoStop}%
\bibitem [{\citenamefont {Nagata}\ and\ \citenamefont
  {Nakamura}(2011)}]{Nagata:2011yf}%
  \BibitemOpen
  \bibfield  {author} {\bibinfo {author} {\bibfnamefont {K.}~\bibnamefont
  {Nagata}}\ and\ \bibinfo {author} {\bibfnamefont {A.}~\bibnamefont
  {Nakamura}},\ }\href {\doibase 10.1103/PhysRevD.83.114507} {\bibfield
  {journal} {\bibinfo  {journal} {Phys.Rev.}\ }\textbf {\bibinfo {volume}
  {D83}},\ \bibinfo {pages} {114507} (\bibinfo {year} {2011})},\ \Eprint
  {http://arxiv.org/abs/1104.2142} {arXiv:1104.2142 [hep-lat]} \BibitemShut
  {NoStop}%
\bibitem [{\citenamefont {Wu}\ and\ \citenamefont {Meng}(2013)}]{Wu:2013bfa}%
  \BibitemOpen
  \bibfield  {author} {\bibinfo {author} {\bibfnamefont {L.-K.}\ \bibnamefont
  {Wu}}\ and\ \bibinfo {author} {\bibfnamefont {X.-F.}\ \bibnamefont {Meng}},\
  }\href@noop {} {\  (\bibinfo {year} {2013})},\ \Eprint
  {http://arxiv.org/abs/1303.0336} {arXiv:1303.0336 [hep-lat]} \BibitemShut
  {NoStop}%
\bibitem [{\citenamefont {Fromm}\ \emph {et~al.}(2012)\citenamefont {Fromm},
  \citenamefont {Langelage}, \citenamefont {Lottini},\ and\ \citenamefont
  {Philipsen}}]{Fromm:2012yg}%
  \BibitemOpen
  \bibfield  {author} {\bibinfo {author} {\bibfnamefont {M.}~\bibnamefont
  {Fromm}}, \bibinfo {author} {\bibfnamefont {J.}~\bibnamefont {Langelage}},
  \bibinfo {author} {\bibfnamefont {S.}~\bibnamefont {Lottini}}, \ and\
  \bibinfo {author} {\bibfnamefont {O.}~\bibnamefont {Philipsen}},\ }\href
  {\doibase 10.1007/JHEP01(2012)042} {\bibfield  {journal} {\bibinfo  {journal}
  {JHEP}\ }\textbf {\bibinfo {volume} {01}},\ \bibinfo {pages} {042} (\bibinfo
  {year} {2012})},\ \Eprint {http://arxiv.org/abs/1111.4953} {arXiv:1111.4953
  [hep-lat]} \BibitemShut {NoStop}%
\bibitem [{\citenamefont {Bluhm}\ and\ \citenamefont
  {Kampfer}(2008)}]{Bluhm:2007cp}%
  \BibitemOpen
  \bibfield  {author} {\bibinfo {author} {\bibfnamefont {M.}~\bibnamefont
  {Bluhm}}\ and\ \bibinfo {author} {\bibfnamefont {B.}~\bibnamefont
  {Kampfer}},\ }\href {\doibase 10.1103/PhysRevD.77.034004} {\bibfield
  {journal} {\bibinfo  {journal} {Phys.Rev.}\ }\textbf {\bibinfo {volume}
  {D77}},\ \bibinfo {pages} {034004} (\bibinfo {year} {2008})},\ \Eprint
  {http://arxiv.org/abs/0711.0590} {arXiv:0711.0590 [hep-ph]} \BibitemShut
  {NoStop}%
\bibitem [{\citenamefont {Braun}\ \emph {et~al.}(2010)\citenamefont {Braun},
  \citenamefont {Gies},\ and\ \citenamefont {Pawlowski}}]{Braun:2007bx}%
  \BibitemOpen
  \bibfield  {author} {\bibinfo {author} {\bibfnamefont {J.}~\bibnamefont
  {Braun}}, \bibinfo {author} {\bibfnamefont {H.}~\bibnamefont {Gies}}, \ and\
  \bibinfo {author} {\bibfnamefont {J.~M.}\ \bibnamefont {Pawlowski}},\ }\href
  {\doibase 10.1016/j.physletb.2010.01.009} {\bibfield  {journal} {\bibinfo
  {journal} {Phys.Lett.}\ }\textbf {\bibinfo {volume} {B684}},\ \bibinfo
  {pages} {262} (\bibinfo {year} {2010})},\ \Eprint
  {http://arxiv.org/abs/0708.2413} {arXiv:0708.2413 [hep-th]} \BibitemShut
  {NoStop}%
\bibitem [{\citenamefont {Sakai}\ \emph
  {et~al.}(2008{\natexlab{a}})\citenamefont {Sakai}, \citenamefont {Kashiwa},
  \citenamefont {Kouno},\ and\ \citenamefont {Yahiro}}]{Sakai:2008um}%
  \BibitemOpen
  \bibfield  {author} {\bibinfo {author} {\bibfnamefont {Y.}~\bibnamefont
  {Sakai}}, \bibinfo {author} {\bibfnamefont {K.}~\bibnamefont {Kashiwa}},
  \bibinfo {author} {\bibfnamefont {H.}~\bibnamefont {Kouno}}, \ and\ \bibinfo
  {author} {\bibfnamefont {M.}~\bibnamefont {Yahiro}},\ }\href {\doibase
  10.1103/PhysRevD.78.036001} {\bibfield  {journal} {\bibinfo  {journal}
  {Phys.Rev.}\ }\textbf {\bibinfo {volume} {D78}},\ \bibinfo {pages} {036001}
  (\bibinfo {year} {2008}{\natexlab{a}})},\ \Eprint
  {http://arxiv.org/abs/0803.1902} {arXiv:0803.1902 [hep-ph]} \BibitemShut
  {NoStop}%
\bibitem [{\citenamefont {Sakai}\ \emph
  {et~al.}(2008{\natexlab{b}})\citenamefont {Sakai}, \citenamefont {Kashiwa},
  \citenamefont {Kouno}, \citenamefont {Matsuzaki},\ and\ \citenamefont
  {Yahiro}}]{Sakai:2008ga}%
  \BibitemOpen
  \bibfield  {author} {\bibinfo {author} {\bibfnamefont {Y.}~\bibnamefont
  {Sakai}}, \bibinfo {author} {\bibfnamefont {K.}~\bibnamefont {Kashiwa}},
  \bibinfo {author} {\bibfnamefont {H.}~\bibnamefont {Kouno}}, \bibinfo
  {author} {\bibfnamefont {M.}~\bibnamefont {Matsuzaki}}, \ and\ \bibinfo
  {author} {\bibfnamefont {M.}~\bibnamefont {Yahiro}},\ }\href {\doibase
  10.1103/PhysRevD.78.076007} {\bibfield  {journal} {\bibinfo  {journal}
  {Phys.Rev.}\ }\textbf {\bibinfo {volume} {D78}},\ \bibinfo {pages} {076007}
  (\bibinfo {year} {2008}{\natexlab{b}})},\ \Eprint
  {http://arxiv.org/abs/0806.4799} {arXiv:0806.4799 [hep-ph]} \BibitemShut
  {NoStop}%
\bibitem [{\citenamefont {Kashiwa}\ \emph
  {et~al.}(2009{\natexlab{a}})\citenamefont {Kashiwa}, \citenamefont
  {Matsuzaki}, \citenamefont {Kouno}, \citenamefont {Sakai},\ and\
  \citenamefont {Yahiro}}]{Kashiwa:2008bq}%
  \BibitemOpen
  \bibfield  {author} {\bibinfo {author} {\bibfnamefont {K.}~\bibnamefont
  {Kashiwa}}, \bibinfo {author} {\bibfnamefont {M.}~\bibnamefont {Matsuzaki}},
  \bibinfo {author} {\bibfnamefont {H.}~\bibnamefont {Kouno}}, \bibinfo
  {author} {\bibfnamefont {Y.}~\bibnamefont {Sakai}}, \ and\ \bibinfo {author}
  {\bibfnamefont {M.}~\bibnamefont {Yahiro}},\ }\href {\doibase
  10.1103/PhysRevD.79.076008} {\bibfield  {journal} {\bibinfo  {journal}
  {Phys.Rev.}\ }\textbf {\bibinfo {volume} {D79}},\ \bibinfo {pages} {076008}
  (\bibinfo {year} {2009}{\natexlab{a}})},\ \Eprint
  {http://arxiv.org/abs/0812.4747} {arXiv:0812.4747 [hep-ph]} \BibitemShut
  {NoStop}%
\bibitem [{\citenamefont {Kouno}\ \emph {et~al.}(2009)\citenamefont {Kouno},
  \citenamefont {Sakai}, \citenamefont {Kashiwa},\ and\ \citenamefont
  {Yahiro}}]{Kouno:2009bm}%
  \BibitemOpen
  \bibfield  {author} {\bibinfo {author} {\bibfnamefont {H.}~\bibnamefont
  {Kouno}}, \bibinfo {author} {\bibfnamefont {Y.}~\bibnamefont {Sakai}},
  \bibinfo {author} {\bibfnamefont {K.}~\bibnamefont {Kashiwa}}, \ and\
  \bibinfo {author} {\bibfnamefont {M.}~\bibnamefont {Yahiro}},\ }\href
  {\doibase 10.1088/0954-3899/36/11/115010} {\bibfield  {journal} {\bibinfo
  {journal} {J.Phys.}\ }\textbf {\bibinfo {volume} {G36}},\ \bibinfo {pages}
  {115010} (\bibinfo {year} {2009})},\ \Eprint {http://arxiv.org/abs/0904.0925}
  {arXiv:0904.0925 [hep-ph]} \BibitemShut {NoStop}%
\bibitem [{\citenamefont {Kashiwa}\ \emph
  {et~al.}(2009{\natexlab{b}})\citenamefont {Kashiwa}, \citenamefont {Kouno},\
  and\ \citenamefont {Yahiro}}]{Kashiwa:2009ki}%
  \BibitemOpen
  \bibfield  {author} {\bibinfo {author} {\bibfnamefont {K.}~\bibnamefont
  {Kashiwa}}, \bibinfo {author} {\bibfnamefont {H.}~\bibnamefont {Kouno}}, \
  and\ \bibinfo {author} {\bibfnamefont {M.}~\bibnamefont {Yahiro}},\ }\href
  {\doibase 10.1103/PhysRevD.80.117901} {\bibfield  {journal} {\bibinfo
  {journal} {Phys.Rev.}\ }\textbf {\bibinfo {volume} {D80}},\ \bibinfo {pages}
  {117901} (\bibinfo {year} {2009}{\natexlab{b}})},\ \Eprint
  {http://arxiv.org/abs/0908.1213} {arXiv:0908.1213 [hep-ph]} \BibitemShut
  {NoStop}%
\bibitem [{\citenamefont {Sakai}\ \emph {et~al.}(2009)\citenamefont {Sakai},
  \citenamefont {Kashiwa}, \citenamefont {Kouno}, \citenamefont {Matsuzaki},\
  and\ \citenamefont {Yahiro}}]{Sakai:2009dv}%
  \BibitemOpen
  \bibfield  {author} {\bibinfo {author} {\bibfnamefont {Y.}~\bibnamefont
  {Sakai}}, \bibinfo {author} {\bibfnamefont {K.}~\bibnamefont {Kashiwa}},
  \bibinfo {author} {\bibfnamefont {H.}~\bibnamefont {Kouno}}, \bibinfo
  {author} {\bibfnamefont {M.}~\bibnamefont {Matsuzaki}}, \ and\ \bibinfo
  {author} {\bibfnamefont {M.}~\bibnamefont {Yahiro}},\ }\href {\doibase
  10.1103/PhysRevD.79.096001} {\bibfield  {journal} {\bibinfo  {journal}
  {Phys.Rev.}\ }\textbf {\bibinfo {volume} {D79}},\ \bibinfo {pages} {096001}
  (\bibinfo {year} {2009})},\ \Eprint {http://arxiv.org/abs/0902.0487}
  {arXiv:0902.0487 [hep-ph]} \BibitemShut {NoStop}%
\bibitem [{\citenamefont {Sakai}\ \emph {et~al.}(2010)\citenamefont {Sakai},
  \citenamefont {Kouno},\ and\ \citenamefont {Yahiro}}]{Sakai:2009vb}%
  \BibitemOpen
  \bibfield  {author} {\bibinfo {author} {\bibfnamefont {Y.}~\bibnamefont
  {Sakai}}, \bibinfo {author} {\bibfnamefont {H.}~\bibnamefont {Kouno}}, \ and\
  \bibinfo {author} {\bibfnamefont {M.}~\bibnamefont {Yahiro}},\ }\href
  {\doibase 10.1088/0954-3899/37/10/105007} {\bibfield  {journal} {\bibinfo
  {journal} {J.Phys.}\ }\textbf {\bibinfo {volume} {G37}},\ \bibinfo {pages}
  {105007} (\bibinfo {year} {2010})},\ \Eprint {http://arxiv.org/abs/0908.3088}
  {arXiv:0908.3088 [hep-ph]} \BibitemShut {NoStop}%
\bibitem [{\citenamefont {Braun}\ \emph {et~al.}(2011)\citenamefont {Braun},
  \citenamefont {Haas}, \citenamefont {Marhauser},\ and\ \citenamefont
  {Pawlowski}}]{Braun:2009gm}%
  \BibitemOpen
  \bibfield  {author} {\bibinfo {author} {\bibfnamefont {J.}~\bibnamefont
  {Braun}}, \bibinfo {author} {\bibfnamefont {L.~M.}\ \bibnamefont {Haas}},
  \bibinfo {author} {\bibfnamefont {F.}~\bibnamefont {Marhauser}}, \ and\
  \bibinfo {author} {\bibfnamefont {J.~M.}\ \bibnamefont {Pawlowski}},\ }\href
  {\doibase 10.1103/PhysRevLett.106.022002} {\bibfield  {journal} {\bibinfo
  {journal} {Phys.Rev.Lett.}\ }\textbf {\bibinfo {volume} {106}},\ \bibinfo
  {pages} {022002} (\bibinfo {year} {2011})},\ \Eprint
  {http://arxiv.org/abs/0908.0008} {arXiv:0908.0008 [hep-ph]} \BibitemShut
  {NoStop}%
\bibitem [{\citenamefont {Nam}(2010)}]{Nam:2009nn}%
  \BibitemOpen
  \bibfield  {author} {\bibinfo {author} {\bibfnamefont {S.-i.}\ \bibnamefont
  {Nam}},\ }\href {\doibase 10.1088/0954-3899/37/7/075002} {\bibfield
  {journal} {\bibinfo  {journal} {J.Phys.}\ }\textbf {\bibinfo {volume}
  {G37}},\ \bibinfo {pages} {075002} (\bibinfo {year} {2010})},\ \Eprint
  {http://arxiv.org/abs/0905.3609} {arXiv:0905.3609 [hep-ph]} \BibitemShut
  {NoStop}%
\bibitem [{\citenamefont {Matsumoto}\ \emph {et~al.}(2011)\citenamefont
  {Matsumoto}, \citenamefont {Kashiwa}, \citenamefont {Kouno}, \citenamefont
  {Oda},\ and\ \citenamefont {Yahiro}}]{Matsumoto:2010vw}%
  \BibitemOpen
  \bibfield  {author} {\bibinfo {author} {\bibfnamefont {T.}~\bibnamefont
  {Matsumoto}}, \bibinfo {author} {\bibfnamefont {K.}~\bibnamefont {Kashiwa}},
  \bibinfo {author} {\bibfnamefont {H.}~\bibnamefont {Kouno}}, \bibinfo
  {author} {\bibfnamefont {K.}~\bibnamefont {Oda}}, \ and\ \bibinfo {author}
  {\bibfnamefont {M.}~\bibnamefont {Yahiro}},\ }\href {\doibase
  10.1016/j.physletb.2010.09.070} {\bibfield  {journal} {\bibinfo  {journal}
  {Phys.Lett.}\ }\textbf {\bibinfo {volume} {B694}},\ \bibinfo {pages} {367}
  (\bibinfo {year} {2011})},\ \Eprint {http://arxiv.org/abs/1004.0592}
  {arXiv:1004.0592 [hep-ph]} \BibitemShut {NoStop}%
\bibitem [{\citenamefont {Kouno}\ \emph {et~al.}(2011)\citenamefont {Kouno},
  \citenamefont {Sakai}, \citenamefont {Sasaki}, \citenamefont {Kashiwa},\ and\
  \citenamefont {Yahiro}}]{Kouno:2011vu}%
  \BibitemOpen
  \bibfield  {author} {\bibinfo {author} {\bibfnamefont {H.}~\bibnamefont
  {Kouno}}, \bibinfo {author} {\bibfnamefont {Y.}~\bibnamefont {Sakai}},
  \bibinfo {author} {\bibfnamefont {T.}~\bibnamefont {Sasaki}}, \bibinfo
  {author} {\bibfnamefont {K.}~\bibnamefont {Kashiwa}}, \ and\ \bibinfo
  {author} {\bibfnamefont {M.}~\bibnamefont {Yahiro}},\ }\href {\doibase
  10.1103/PhysRevD.83.076009} {\bibfield  {journal} {\bibinfo  {journal}
  {Phys.Rev.}\ }\textbf {\bibinfo {volume} {D83}},\ \bibinfo {pages} {076009}
  (\bibinfo {year} {2011})},\ \Eprint {http://arxiv.org/abs/1101.5746}
  {arXiv:1101.5746 [hep-ph]} \BibitemShut {NoStop}%
\bibitem [{\citenamefont {Kouno}\ \emph {et~al.}(2012)\citenamefont {Kouno},
  \citenamefont {Kishikawa}, \citenamefont {Sasaki}, \citenamefont {Sakai},\
  and\ \citenamefont {Yahiro}}]{Kouno:2011zu}%
  \BibitemOpen
  \bibfield  {author} {\bibinfo {author} {\bibfnamefont {H.}~\bibnamefont
  {Kouno}}, \bibinfo {author} {\bibfnamefont {M.}~\bibnamefont {Kishikawa}},
  \bibinfo {author} {\bibfnamefont {T.}~\bibnamefont {Sasaki}}, \bibinfo
  {author} {\bibfnamefont {Y.}~\bibnamefont {Sakai}}, \ and\ \bibinfo {author}
  {\bibfnamefont {M.}~\bibnamefont {Yahiro}},\ }\href {\doibase
  10.1103/PhysRevD.85.016001} {\bibfield  {journal} {\bibinfo  {journal}
  {Phys.Rev.}\ }\textbf {\bibinfo {volume} {D85}},\ \bibinfo {pages} {016001}
  (\bibinfo {year} {2012})},\ \Eprint {http://arxiv.org/abs/1110.5187}
  {arXiv:1110.5187 [hep-ph]} \BibitemShut {NoStop}%
\bibitem [{\citenamefont {Sasaki}\ \emph {et~al.}(2011)\citenamefont {Sasaki},
  \citenamefont {Sakai}, \citenamefont {Kouno},\ and\ \citenamefont
  {Yahiro}}]{Sasaki:2011}%
  \BibitemOpen
  \bibfield  {author} {\bibinfo {author} {\bibfnamefont {T.}~\bibnamefont
  {Sasaki}}, \bibinfo {author} {\bibfnamefont {Y.}~\bibnamefont {Sakai}},
  \bibinfo {author} {\bibfnamefont {H.}~\bibnamefont {Kouno}}, \ and\ \bibinfo
  {author} {\bibfnamefont {M.}~\bibnamefont {Yahiro}},\ }\href {\doibase
  10.1103/PhysRevD.84.091901} {\bibfield  {journal} {\bibinfo  {journal} {Phys.
  Rev. D}\ }\textbf {\bibinfo {volume} {84}},\ \bibinfo {pages} {091901}
  (\bibinfo {year} {2011})},\ \Eprint {http://arxiv.org/abs/1105.3959}
  {arXiv:1105.3959 [hep-hp]} \BibitemShut {NoStop}%
\bibitem [{\citenamefont {Morita}\ \emph
  {et~al.}(2011{\natexlab{a}})\citenamefont {Morita}, \citenamefont {Skokov},
  \citenamefont {Friman},\ and\ \citenamefont {Redlich}}]{Morita:2011eu}%
  \BibitemOpen
  \bibfield  {author} {\bibinfo {author} {\bibfnamefont {K.}~\bibnamefont
  {Morita}}, \bibinfo {author} {\bibfnamefont {V.}~\bibnamefont {Skokov}},
  \bibinfo {author} {\bibfnamefont {B.}~\bibnamefont {Friman}}, \ and\ \bibinfo
  {author} {\bibfnamefont {K.}~\bibnamefont {Redlich}},\ }\href {\doibase
  10.1103/PhysRevD.84.076009} {\bibfield  {journal} {\bibinfo  {journal}
  {Phys.Rev.}\ }\textbf {\bibinfo {volume} {D84}},\ \bibinfo {pages} {076009}
  (\bibinfo {year} {2011}{\natexlab{a}})},\ \Eprint
  {http://arxiv.org/abs/1107.2273} {arXiv:1107.2273 [hep-ph]} \BibitemShut
  {NoStop}%
\bibitem [{\citenamefont {Morita}\ \emph
  {et~al.}(2011{\natexlab{b}})\citenamefont {Morita}, \citenamefont {Skokov},
  \citenamefont {Friman},\ and\ \citenamefont {Redlich}}]{Morita:2011jva}%
  \BibitemOpen
  \bibfield  {author} {\bibinfo {author} {\bibfnamefont {K.}~\bibnamefont
  {Morita}}, \bibinfo {author} {\bibfnamefont {V.}~\bibnamefont {Skokov}},
  \bibinfo {author} {\bibfnamefont {B.}~\bibnamefont {Friman}}, \ and\ \bibinfo
  {author} {\bibfnamefont {K.}~\bibnamefont {Redlich}},\ }\href {\doibase
  10.1103/PhysRevD.84.074020} {\bibfield  {journal} {\bibinfo  {journal}
  {Phys.Rev.}\ }\textbf {\bibinfo {volume} {D84}},\ \bibinfo {pages} {074020}
  (\bibinfo {year} {2011}{\natexlab{b}})},\ \Eprint
  {http://arxiv.org/abs/1108.0735} {arXiv:1108.0735 [hep-ph]} \BibitemShut
  {NoStop}%
\bibitem [{\citenamefont {Pagura}\ \emph {et~al.}(2012)\citenamefont {Pagura},
  \citenamefont {Gomez~Dumm},\ and\ \citenamefont {Scoccola}}]{Pagura:2011rt}%
  \BibitemOpen
  \bibfield  {author} {\bibinfo {author} {\bibfnamefont {V.}~\bibnamefont
  {Pagura}}, \bibinfo {author} {\bibfnamefont {D.}~\bibnamefont {Gomez~Dumm}},
  \ and\ \bibinfo {author} {\bibfnamefont {N.}~\bibnamefont {Scoccola}},\
  }\href {\doibase 10.1016/j.physletb.2011.11.064} {\bibfield  {journal}
  {\bibinfo  {journal} {Phys.Lett.}\ }\textbf {\bibinfo {volume} {B707}},\
  \bibinfo {pages} {76} (\bibinfo {year} {2012})},\ \Eprint
  {http://arxiv.org/abs/1105.1739} {arXiv:1105.1739 [hep-ph]} \BibitemShut
  {NoStop}%
\bibitem [{\citenamefont {Kashiwa}\ \emph {et~al.}(2013)\citenamefont
  {Kashiwa}, \citenamefont {Sasaki}, \citenamefont {Kouno},\ and\ \citenamefont
  {Yahiro}}]{Kashiwa:2012xm}%
  \BibitemOpen
  \bibfield  {author} {\bibinfo {author} {\bibfnamefont {K.}~\bibnamefont
  {Kashiwa}}, \bibinfo {author} {\bibfnamefont {T.}~\bibnamefont {Sasaki}},
  \bibinfo {author} {\bibfnamefont {H.}~\bibnamefont {Kouno}}, \ and\ \bibinfo
  {author} {\bibfnamefont {M.}~\bibnamefont {Yahiro}},\ }\href {\doibase
  10.1103/PhysRevD.87.016015} {\bibfield  {journal} {\bibinfo  {journal}
  {Phys.Rev.}\ }\textbf {\bibinfo {volume} {D87}},\ \bibinfo {pages} {016015}
  (\bibinfo {year} {2013})},\ \Eprint {http://arxiv.org/abs/1208.2283}
  {arXiv:1208.2283 [hep-ph]} \BibitemShut {NoStop}%
\bibitem [{\citenamefont {Fister}\ and\ \citenamefont
  {Pawlowski}(2013)}]{Fister:2013bh}%
  \BibitemOpen
  \bibfield  {author} {\bibinfo {author} {\bibfnamefont {L.}~\bibnamefont
  {Fister}}\ and\ \bibinfo {author} {\bibfnamefont {J.~M.}\ \bibnamefont
  {Pawlowski}},\ }\href@noop {} {\  (\bibinfo {year} {2013})},\ \Eprint
  {http://arxiv.org/abs/1301.4163} {arXiv:1301.4163 [hep-ph]} \BibitemShut
  {NoStop}%
\bibitem [{\citenamefont {Aarts}\ \emph {et~al.}(2010)\citenamefont {Aarts},
  \citenamefont {Kumar},\ and\ \citenamefont {Rafferty}}]{Aarts:2010ky}%
  \BibitemOpen
  \bibfield  {author} {\bibinfo {author} {\bibfnamefont {G.}~\bibnamefont
  {Aarts}}, \bibinfo {author} {\bibfnamefont {S.~P.}\ \bibnamefont {Kumar}}, \
  and\ \bibinfo {author} {\bibfnamefont {J.}~\bibnamefont {Rafferty}},\ }\href
  {\doibase 10.1007/JHEP07(2010)056} {\bibfield  {journal} {\bibinfo  {journal}
  {JHEP}\ }\textbf {\bibinfo {volume} {1007}},\ \bibinfo {pages} {056}
  (\bibinfo {year} {2010})},\ \Eprint {http://arxiv.org/abs/1005.2947}
  {arXiv:1005.2947 [hep-th]} \BibitemShut {NoStop}%
\bibitem [{\citenamefont {Rafferty}(2011)}]{Rafferty:2011hd}%
  \BibitemOpen
  \bibfield  {author} {\bibinfo {author} {\bibfnamefont {J.}~\bibnamefont
  {Rafferty}},\ }\href {\doibase 10.1007/JHEP09(2011)087} {\bibfield  {journal}
  {\bibinfo  {journal} {JHEP}\ }\textbf {\bibinfo {volume} {1109}},\ \bibinfo
  {pages} {087} (\bibinfo {year} {2011})},\ \Eprint
  {http://arxiv.org/abs/1103.2315} {arXiv:1103.2315 [hep-th]} \BibitemShut
  {NoStop}%
\bibitem [{\citenamefont {'t~Hooft}(1979)}]{'tHooft:1979uj}%
  \BibitemOpen
  \bibfield  {author} {\bibinfo {author} {\bibfnamefont {G.}~\bibnamefont
  {'t~Hooft}},\ }\href {\doibase 10.1016/0550-3213(79)90595-9} {\bibfield
  {journal} {\bibinfo  {journal} {Nucl.Phys.}\ }\textbf {\bibinfo {volume}
  {B153}},\ \bibinfo {pages} {141} (\bibinfo {year} {1979})}\BibitemShut
  {NoStop}%
\bibitem [{\citenamefont {Korthals-Altes}\ \emph {et~al.}(1999)\citenamefont
  {Korthals-Altes}, \citenamefont {Kovner},\ and\ \citenamefont
  {Stephanov}}]{KorthalsAltes:1999xb}%
  \BibitemOpen
  \bibfield  {author} {\bibinfo {author} {\bibfnamefont {C.}~\bibnamefont
  {Korthals-Altes}}, \bibinfo {author} {\bibfnamefont {A.}~\bibnamefont
  {Kovner}}, \ and\ \bibinfo {author} {\bibfnamefont {M.~A.}\ \bibnamefont
  {Stephanov}},\ }\href {\doibase 10.1016/S0370-2693(99)01242-3} {\bibfield
  {journal} {\bibinfo  {journal} {Phys.Lett.}\ }\textbf {\bibinfo {volume}
  {B469}},\ \bibinfo {pages} {205} (\bibinfo {year} {1999})},\ \Eprint
  {http://arxiv.org/abs/hep-ph/9909516} {arXiv:hep-ph/9909516 [hep-ph]}
  \BibitemShut {NoStop}%
\bibitem [{\citenamefont {Korthals-Altes}\ and\ \citenamefont
  {Kovner}(2000)}]{KorthalsAltes:2000gs}%
  \BibitemOpen
  \bibfield  {author} {\bibinfo {author} {\bibfnamefont {C.}~\bibnamefont
  {Korthals-Altes}}\ and\ \bibinfo {author} {\bibfnamefont {A.}~\bibnamefont
  {Kovner}},\ }\href {\doibase 10.1103/PhysRevD.62.096008} {\bibfield
  {journal} {\bibinfo  {journal} {Phys.Rev.}\ }\textbf {\bibinfo {volume}
  {D62}},\ \bibinfo {pages} {096008} (\bibinfo {year} {2000})},\ \Eprint
  {http://arxiv.org/abs/hep-ph/0004052} {arXiv:hep-ph/0004052 [hep-ph]}
  \BibitemShut {NoStop}%
\bibitem [{\citenamefont {de~Forcrand}\ and\ \citenamefont {von
  Smekal}(2002)}]{deForcrand:2001nd}%
  \BibitemOpen
  \bibfield  {author} {\bibinfo {author} {\bibfnamefont {P.}~\bibnamefont
  {de~Forcrand}}\ and\ \bibinfo {author} {\bibfnamefont {L.}~\bibnamefont {von
  Smekal}},\ }\href {\doibase 10.1103/PhysRevD.66.011504} {\bibfield  {journal}
  {\bibinfo  {journal} {Phys.Rev.}\ }\textbf {\bibinfo {volume} {D66}},\
  \bibinfo {pages} {011504} (\bibinfo {year} {2002})},\ \Eprint
  {http://arxiv.org/abs/hep-lat/0107018} {arXiv:hep-lat/0107018 [hep-lat]}
  \BibitemShut {NoStop}%
\bibitem [{\citenamefont {de~Forcrand}\ \emph {et~al.}(2001)\citenamefont
  {de~Forcrand}, \citenamefont {D'Elia},\ and\ \citenamefont
  {Pepe}}]{deForcrand:2000fi}%
  \BibitemOpen
  \bibfield  {author} {\bibinfo {author} {\bibfnamefont {P.}~\bibnamefont
  {de~Forcrand}}, \bibinfo {author} {\bibfnamefont {M.}~\bibnamefont {D'Elia}},
  \ and\ \bibinfo {author} {\bibfnamefont {M.}~\bibnamefont {Pepe}},\ }\href
  {\doibase 10.1103/PhysRevLett.86.1438} {\bibfield  {journal} {\bibinfo
  {journal} {Phys.Rev.Lett.}\ }\textbf {\bibinfo {volume} {86}},\ \bibinfo
  {pages} {1438} (\bibinfo {year} {2001})},\ \Eprint
  {http://arxiv.org/abs/hep-lat/0007034} {arXiv:hep-lat/0007034 [hep-lat]}
  \BibitemShut {NoStop}%
\bibitem [{\citenamefont {de~Forcrand}\ and\ \citenamefont
  {Noth}(2005)}]{deForcrand:2005pb}%
  \BibitemOpen
  \bibfield  {author} {\bibinfo {author} {\bibfnamefont {P.}~\bibnamefont
  {de~Forcrand}}\ and\ \bibinfo {author} {\bibfnamefont {D.}~\bibnamefont
  {Noth}},\ }\href {\doibase 10.1103/PhysRevD.72.114501} {\bibfield  {journal}
  {\bibinfo  {journal} {Phys.Rev.}\ }\textbf {\bibinfo {volume} {D72}},\
  \bibinfo {pages} {114501} (\bibinfo {year} {2005})},\ \Eprint
  {http://arxiv.org/abs/hep-lat/0506005} {arXiv:hep-lat/0506005 [hep-lat]}
  \BibitemShut {NoStop}%
\bibitem [{\citenamefont {Lucini}\ and\ \citenamefont
  {Panero}(2012)}]{Lucini:2012gg}%
  \BibitemOpen
  \bibfield  {author} {\bibinfo {author} {\bibfnamefont {B.}~\bibnamefont
  {Lucini}}\ and\ \bibinfo {author} {\bibfnamefont {M.}~\bibnamefont
  {Panero}},\ }\href@noop {} {\  (\bibinfo {year} {2012})},\ \Eprint
  {http://arxiv.org/abs/1210.4997} {arXiv:1210.4997 [hep-th]} \BibitemShut
  {NoStop}%
\bibitem [{\citenamefont {Bhattacharya}\ \emph {et~al.}(1991)\citenamefont
  {Bhattacharya}, \citenamefont {Gocksch}, \citenamefont {Korthals~Altes},\
  and\ \citenamefont {Pisarski}}]{Bhattacharya:1990hk}%
  \BibitemOpen
  \bibfield  {author} {\bibinfo {author} {\bibfnamefont {T.}~\bibnamefont
  {Bhattacharya}}, \bibinfo {author} {\bibfnamefont {A.}~\bibnamefont
  {Gocksch}}, \bibinfo {author} {\bibfnamefont {C.}~\bibnamefont
  {Korthals~Altes}}, \ and\ \bibinfo {author} {\bibfnamefont {R.~D.}\
  \bibnamefont {Pisarski}},\ }\href {\doibase 10.1103/PhysRevLett.66.998}
  {\bibfield  {journal} {\bibinfo  {journal} {Phys.Rev.Lett.}\ }\textbf
  {\bibinfo {volume} {66}},\ \bibinfo {pages} {998} (\bibinfo {year}
  {1991})}\BibitemShut {NoStop}%
\bibitem [{\citenamefont {Bhattacharya}\ \emph {et~al.}(1992)\citenamefont
  {Bhattacharya}, \citenamefont {Gocksch}, \citenamefont {Korthals~Altes},\
  and\ \citenamefont {Pisarski}}]{Bhattacharya:1992qb}%
  \BibitemOpen
  \bibfield  {author} {\bibinfo {author} {\bibfnamefont {T.}~\bibnamefont
  {Bhattacharya}}, \bibinfo {author} {\bibfnamefont {A.}~\bibnamefont
  {Gocksch}}, \bibinfo {author} {\bibfnamefont {C.}~\bibnamefont
  {Korthals~Altes}}, \ and\ \bibinfo {author} {\bibfnamefont {R.~D.}\
  \bibnamefont {Pisarski}},\ }\href {\doibase 10.1016/0550-3213(92)90086-Q}
  {\bibfield  {journal} {\bibinfo  {journal} {Nucl.Phys.}\ }\textbf {\bibinfo
  {volume} {B383}},\ \bibinfo {pages} {497} (\bibinfo {year} {1992})},\ \Eprint
  {http://arxiv.org/abs/hep-ph/9205231} {arXiv:hep-ph/9205231 [hep-ph]}
  \BibitemShut {NoStop}%
\bibitem [{\citenamefont {Giovannangeli}\ and\ \citenamefont
  {Korthals~Altes}(2005{\natexlab{a}})}]{Giovannangeli:2002uv}%
  \BibitemOpen
  \bibfield  {author} {\bibinfo {author} {\bibfnamefont {P.}~\bibnamefont
  {Giovannangeli}}\ and\ \bibinfo {author} {\bibfnamefont {C.}~\bibnamefont
  {Korthals~Altes}},\ }\href {\doibase 10.1016/j.nuclphysb.2005.05.010}
  {\bibfield  {journal} {\bibinfo  {journal} {Nucl.Phys.}\ }\textbf {\bibinfo
  {volume} {B721}},\ \bibinfo {pages} {1} (\bibinfo {year}
  {2005}{\natexlab{a}})},\ \Eprint {http://arxiv.org/abs/hep-ph/0212298}
  {arXiv:hep-ph/0212298 [hep-ph]} \BibitemShut {NoStop}%
\bibitem [{\citenamefont {Giovannangeli}\ and\ \citenamefont
  {Korthals~Altes}(2005{\natexlab{b}})}]{Giovannangeli:2004sg}%
  \BibitemOpen
  \bibfield  {author} {\bibinfo {author} {\bibfnamefont {P.}~\bibnamefont
  {Giovannangeli}}\ and\ \bibinfo {author} {\bibfnamefont {C.}~\bibnamefont
  {Korthals~Altes}},\ }\href {\doibase 10.1016/j.nuclphysb.2005.03.024}
  {\bibfield  {journal} {\bibinfo  {journal} {Nucl.Phys.}\ }\textbf {\bibinfo
  {volume} {B721}},\ \bibinfo {pages} {25} (\bibinfo {year}
  {2005}{\natexlab{b}})},\ \Eprint {http://arxiv.org/abs/hep-ph/0412322}
  {arXiv:hep-ph/0412322 [hep-ph]} \BibitemShut {NoStop}%
\bibitem [{\citenamefont {Dumitru}\ \emph {et~al.}(2011)\citenamefont
  {Dumitru}, \citenamefont {Guo}, \citenamefont {Hidaka}, \citenamefont
  {Altes},\ and\ \citenamefont {Pisarski}}]{Dumitru:2011}%
  \BibitemOpen
  \bibfield  {author} {\bibinfo {author} {\bibfnamefont {A.}~\bibnamefont
  {Dumitru}}, \bibinfo {author} {\bibfnamefont {Y.}~\bibnamefont {Guo}},
  \bibinfo {author} {\bibfnamefont {Y.}~\bibnamefont {Hidaka}}, \bibinfo
  {author} {\bibfnamefont {C.~P.~K.}\ \bibnamefont {Altes}}, \ and\ \bibinfo
  {author} {\bibfnamefont {R.~D.}\ \bibnamefont {Pisarski}},\ }\href {\doibase
  10.1103/PhysRevD.83.034022} {\bibfield  {journal} {\bibinfo  {journal} {Phys.
  Rev. D}\ }\textbf {\bibinfo {volume} {83}},\ \bibinfo {pages} {034022}
  (\bibinfo {year} {2011})},\ \Eprint {http://arxiv.org/abs/1011.3820}
  {arXiv:1011.3820 [hep-ph]} \BibitemShut {NoStop}%
\bibitem [{\citenamefont {Dumitru}\ \emph {et~al.}(2012)\citenamefont
  {Dumitru}, \citenamefont {Guo}, \citenamefont {Hidaka}, \citenamefont
  {Altes},\ and\ \citenamefont {Pisarski}}]{Dumitru:2012fw}%
  \BibitemOpen
  \bibfield  {author} {\bibinfo {author} {\bibfnamefont {A.}~\bibnamefont
  {Dumitru}}, \bibinfo {author} {\bibfnamefont {Y.}~\bibnamefont {Guo}},
  \bibinfo {author} {\bibfnamefont {Y.}~\bibnamefont {Hidaka}}, \bibinfo
  {author} {\bibfnamefont {C.~P.~K.}\ \bibnamefont {Altes}}, \ and\ \bibinfo
  {author} {\bibfnamefont {R.~D.}\ \bibnamefont {Pisarski}},\ }\href {\doibase
  10.1103/PhysRevD.86.105017} {\bibfield  {journal} {\bibinfo  {journal}
  {Phys.Rev.}\ }\textbf {\bibinfo {volume} {D86}},\ \bibinfo {pages} {105017}
  (\bibinfo {year} {2012})},\ \Eprint {http://arxiv.org/abs/1205.0137}
  {arXiv:1205.0137 [hep-ph]} \BibitemShut {NoStop}%
\bibitem [{\citenamefont {Pisarski}\ and\ \citenamefont
  {Skokov}(2012)}]{Pisarski:2012bj}%
  \BibitemOpen
  \bibfield  {author} {\bibinfo {author} {\bibfnamefont {R.~D.}\ \bibnamefont
  {Pisarski}}\ and\ \bibinfo {author} {\bibfnamefont {V.~V.}\ \bibnamefont
  {Skokov}},\ }\href {\doibase 10.1103/PhysRevD.86.081701} {\bibfield
  {journal} {\bibinfo  {journal} {Phys.Rev.}\ }\textbf {\bibinfo {volume}
  {D86}},\ \bibinfo {pages} {081701} (\bibinfo {year} {2012})},\ \Eprint
  {http://arxiv.org/abs/1206.1329} {arXiv:1206.1329 [hep-th]} \BibitemShut
  {NoStop}%
\bibitem [{\citenamefont {Kashiwa}\ \emph {et~al.}(2012)\citenamefont
  {Kashiwa}, \citenamefont {Pisarski},\ and\ \citenamefont
  {Skokov}}]{Kashiwa:2012va}%
  \BibitemOpen
  \bibfield  {author} {\bibinfo {author} {\bibfnamefont {K.}~\bibnamefont
  {Kashiwa}}, \bibinfo {author} {\bibfnamefont {R.~D.}\ \bibnamefont
  {Pisarski}}, \ and\ \bibinfo {author} {\bibfnamefont {V.~V.}\ \bibnamefont
  {Skokov}},\ }\href {\doibase 10.1103/PhysRevD.85.114029} {\bibfield
  {journal} {\bibinfo  {journal} {Phys. Rev. D}\ }\textbf {\bibinfo {volume}
  {85}},\ \bibinfo {pages} {114029} (\bibinfo {year} {2012})},\ \Eprint
  {http://arxiv.org/abs/1205.0545} {arXiv:1205.0545 [hep-ph]} \BibitemShut
  {NoStop}%
\bibitem [{\citenamefont {Weiss}(1981)}]{Weiss:1981}%
  \BibitemOpen
  \bibfield  {author} {\bibinfo {author} {\bibfnamefont {N.}~\bibnamefont
  {Weiss}},\ }\href {\doibase 10.1103/PhysRevD.24.475} {\bibfield  {journal}
  {\bibinfo  {journal} {Phys. Rev. D}\ }\textbf {\bibinfo {volume} {24}},\
  \bibinfo {pages} {475} (\bibinfo {year} {1981})}\BibitemShut {NoStop}%
\bibitem [{\citenamefont {Gross}\ \emph {et~al.}(1981)\citenamefont {Gross},
  \citenamefont {Pisarski},\ and\ \citenamefont {Yaffe}}]{Gross:1981br}%
  \BibitemOpen
  \bibfield  {author} {\bibinfo {author} {\bibfnamefont {D.~J.}\ \bibnamefont
  {Gross}}, \bibinfo {author} {\bibfnamefont {R.~D.}\ \bibnamefont {Pisarski}},
  \ and\ \bibinfo {author} {\bibfnamefont {L.~G.}\ \bibnamefont {Yaffe}},\
  }\href {\doibase 10.1103/RevModPhys.53.43} {\bibfield  {journal} {\bibinfo
  {journal} {Rev.Mod.Phys.}\ }\textbf {\bibinfo {volume} {53}},\ \bibinfo
  {pages} {43} (\bibinfo {year} {1981})}\BibitemShut {NoStop}%
\bibitem [{\citenamefont {Belyaev}\ \emph {et~al.}(1992)\citenamefont
  {Belyaev}, \citenamefont {Kogan}, \citenamefont {Semenoff},\ and\
  \citenamefont {Weiss}}]{Belyaev:1991np}%
  \BibitemOpen
  \bibfield  {author} {\bibinfo {author} {\bibfnamefont {V.}~\bibnamefont
  {Belyaev}}, \bibinfo {author} {\bibfnamefont {I.~I.}\ \bibnamefont {Kogan}},
  \bibinfo {author} {\bibfnamefont {G.}~\bibnamefont {Semenoff}}, \ and\
  \bibinfo {author} {\bibfnamefont {N.}~\bibnamefont {Weiss}},\ }\href
  {\doibase 10.1016/0370-2693(92)90754-R} {\bibfield  {journal} {\bibinfo
  {journal} {Phys.Lett.}\ }\textbf {\bibinfo {volume} {B277}},\ \bibinfo
  {pages} {331} (\bibinfo {year} {1992})}\BibitemShut {NoStop}%
\bibitem [{\citenamefont {Smilga}(1994)}]{Smilga:1993vb}%
  \BibitemOpen
  \bibfield  {author} {\bibinfo {author} {\bibfnamefont {A.~V.}\ \bibnamefont
  {Smilga}},\ }\href {\doibase 10.1006/aphy.1994.1073} {\bibfield  {journal}
  {\bibinfo  {journal} {Annals Phys.}\ }\textbf {\bibinfo {volume} {234}},\
  \bibinfo {pages} {1} (\bibinfo {year} {1994})}\BibitemShut {NoStop}%
\bibitem [{\citenamefont {Fukushima}\ and\ \citenamefont
  {Kashiwa}(2012)}]{Fukushima:2012}%
  \BibitemOpen
  \bibfield  {author} {\bibinfo {author} {\bibfnamefont {K.}~\bibnamefont
  {Fukushima}}\ and\ \bibinfo {author} {\bibfnamefont {K.}~\bibnamefont
  {Kashiwa}},\ }\href@noop {} {\  (\bibinfo {year} {2012})},\ \Eprint
  {http://arxiv.org/abs/1206.0685} {arXiv:1206.0685 [hep-ph]} \BibitemShut
  {NoStop}%
\end{thebibliography}%

\end{document}